\documentclass[letterpaper,twocolumn,prl,aps,superscriptaddress,floatfix]{revtex4}
\usepackage{mathptmx}

\usepackage[latin9]{inputenc}
\usepackage{amsmath}
\usepackage{amssymb}
\usepackage{graphicx}
\usepackage{esint}
\usepackage{array}
\pdfpageheight\paperheight
\pdfpagewidth\paperwidth

\providecommand{\tabularnewline}{\\}

\usepackage{xcolor}\usepackage{soul}

\newcommand{\ket}[1]{\ensuremath{\left|#1\right\rangle}}
\definecolor{blue}{rgb}{0,0,1}
\definecolor{red}{rgb}{1,0,0}
\definecolor{green}{rgb}{0,1,0}

\makeatother

\begin{document}

\title{Perfect quantum state transfer in a superconducting qubit chain\\ with parametrically tunable couplings}
\author{X. Li}
\thanks{These two authors contributed equally to this work.}
\affiliation{Center for Quantum Information, Institute for Interdisciplinary Information Sciences, Tsinghua University, Beijing 100084, China}

\author{Y. Ma}
\thanks{These two authors contributed equally to this work.}
\affiliation{Center for Quantum Information, Institute for Interdisciplinary Information Sciences, Tsinghua University, Beijing 100084, China}

\author{J. Han}
\affiliation{Center for Quantum Information, Institute for Interdisciplinary Information Sciences, Tsinghua University, Beijing 100084, China}

\author{Tao Chen}
\affiliation{Guangdong Provincial Key Laboratory of Quantum Engineering
and Quantum Materials, and School of Physics\\ and Telecommunication Engineering,
South China Normal University, Guangzhou 510006, China}

\author{Y. Xu}
\affiliation{Center for Quantum Information, Institute for Interdisciplinary Information Sciences, Tsinghua University, Beijing 100084, China}

\author{W. Cai}
\affiliation{Center for Quantum Information, Institute for Interdisciplinary Information Sciences, Tsinghua University, Beijing 100084, China}

\author{H. Wang}
\affiliation{Center for Quantum Information, Institute for Interdisciplinary Information Sciences, Tsinghua University, Beijing 100084, China}

\author{Y. P. Song}
\affiliation{Center for Quantum Information, Institute for Interdisciplinary Information Sciences, Tsinghua University, Beijing 100084, China}

\author{Zheng-Yuan Xue} \email{zyxue83@163.com}
\affiliation{Guangdong Provincial Key Laboratory of Quantum Engineering
and Quantum Materials, and School of Physics\\ and Telecommunication Engineering,
South China Normal University, Guangzhou 510006, China}

\author{Zhang-qi Yin}\email{yinzhangqi@tsinghua.edu.cn}

\author{Luyan Sun}\email{luyansun@tsinghua.edu.cn}
\affiliation{Center for Quantum Information, Institute for Interdisciplinary Information Sciences, Tsinghua University, Beijing 100084, China}

\begin{abstract}

Faithfully transferring quantum state is essential for quantum information processing. Here, we demonstrate a fast (in 84~ns) and high-fidelity (99.2\%) transfer of arbitrary quantum states in a chain of four superconducting qubits with nearest-neighbor coupling. This transfer relies on full control of the effective couplings between neighboring qubits, which is realized only by parametrically modulating the qubits without increasing circuit complexity. Once the couplings between qubits fulfill specific ratio, a perfect quantum state transfer can be achieved in a single step, therefore robust to noise and accumulation of experimental errors. This quantum state transfer can be extended to a larger qubit chain and thus adds a desirable tool for future quantum information processing. The demonstrated flexibility of the coupling tunability is suitable for quantum simulation of many-body physics which requires different configurations of qubit couplings.

\end{abstract}
%\date{\today}

\maketitle
\section{Introduction}
High-fidelity quantum state transfer (QST) from one place to another is important for both quantum communication and scalable quantum computation~\cite{Divincenzo}. Long-distance QST is an essential element for a quantum network~\cite{Kimble2008}, which requires high-efficiency interfaces for the transfer of stationary quantum states to flying photonic qubits~\cite{Cirac1997,Serafini2006,Yin2007,Axline2018,Kurpiers2017}. On the other hand, short-distance QST is important for on-chip quantum information processing, such as communication among quantum processors and writing (reading) quantum information into (out of) quantum memories~\cite{Nielsen}. To achieve short-distance QST, which is our main focus, previously reported methods include physically moving the qubits holding quantum information to other target sites~\cite{Kielpinski2002}. However, solid-state qubits are usually static and coupled with each other. Therefore, it is desired to realize QST in such systems solely through the interactions between qubits. This goal could be achieved by sequential swap operations between nearest neighbors through active control of the qubits, but this method will result in accumulation of individual operation errors.

Alternatively, QST in a qubit chain can be realized in a single step~\cite{Bose2003,qst2004,Romito2005,Yung2005,Shi2005,Bose2007,Franco2008,Yao2011}, which eliminates the error-accumulation problem. In particular, perfect QST in an arbitrarily long chain was proposed in Ref.~\onlinecite{qst2004} without active qubit control when the coupling strengths between the qubits fulfill specific conditions. Since then, the perfect-QST protocol~\cite{qst2004} was simulated in nuclear magnetic resonance~\cite{Zhang2005}, and was demonstrated in optical systems~\cite{Perez2013,Chapman2016} with a fixed number of waveguides. However, it is challenging to realize this protocol in solid-state quantum systems because it is typically difficult to precisely preset the couplings among qubits for a fixed chain. It is even more difficult to realize  perfect QST with variable lengths in the chain. Therefore, to meet the perfect-QST condition~\cite{qst2004}, tunable coupling between qubits is necessary. Moreover, this tunability is also of particular importance for realizing two-qubit quantum gates~\cite{Chow2011,Poletto2012,DiCarlo2009,Kelly2015}. However, the coupling tunability usually comes at the cost of additional decoherence or circuit complexity~\cite{Liu2006,Niskanen2007,McKay2016,Naik2017,Lu2017,chen2017,Roushan2017,Neill2018}.

In this work, we adopt a method of parametric modulation of qubit frequencies to realize tunable qubit coupling strengths~\cite{Zhou2009,Strand2013,Liu2014,Wu2016,Caldwell2017,Reagor2018} and experimentally demonstrate perfect QST in a chain of four coupled superconducting transmon qubits. In this method, the modulating fields provide control over qubit-qubit interactions without relying on extra coupling elements, and thus the circuit complexity remains the simplest. In our experiment, we first realize a large-range tunable coupling between two nearest-neighbor qubits by parametrically modulating only one qubit. We next verify the coupling-strength tunability for a chain with multiple modulations, where the qubits in the middle are affected by multiple fields, therefore expanding the parametric tunability toolbox. Finally, we apply this technique to a chain with four qubits and experimentally realize QST in $84$~ns with a fidelity of $99.2\%$, characterized by quantum process tomography. Our experiment is thus in sharp contrast to the previous demonstrations in optical systems. The demonstrated flexibility of the coupling tunability can be further applied to quantum simulation experiments~\cite{Houck2012,Georgescu2014,Kyriienko2018}, which require different configurations of qubit coupling strengths.

\begin{figure}
\includegraphics{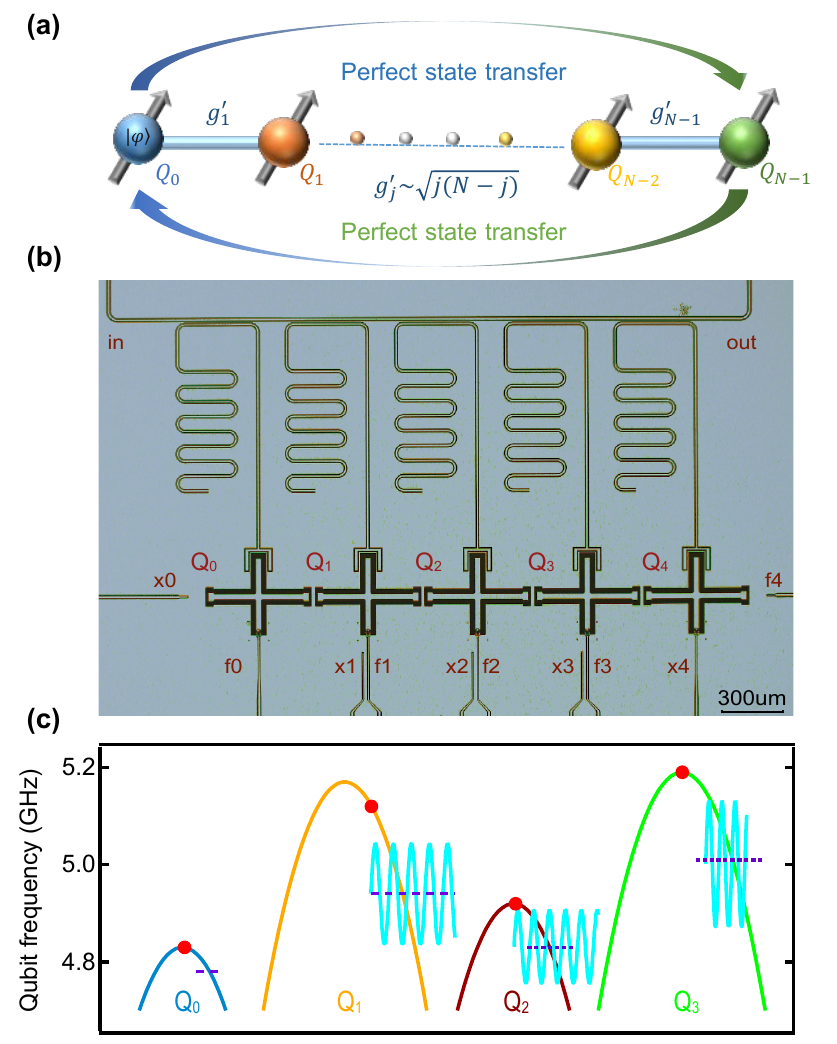}
\caption{ (a) Perfect QST. The quantum state is transferred from the first qubit $Q_0$ to the last qubit $Q_{N-1}$ in a chain when the couplings between neighboring qubits satisfy specific ratio.  (b) The five-qubit chain sample. Five cross-shaped transmon qubits (Xmons, $Q_0-Q_4$) arranged in a linear array. Each qubit has independent $XY$ and $Z$ control (labeled as $``x"$ and $``f"$ respectively), and is coupled to a separate $\lambda/4$ resonator for simultaneous and individual read-out.  (c) The operating regime for QST from $Q_0$ to $Q_3$ with $Q_4$ being decoupled. Both state preparation and measurement of each qubit are performed at the red points (the idle points), which are the maximum-frequency spots (sweet spots) with the best coherence times for $Q_0, Q_2$, and $Q_3$. $Q_1$ is biased about 60~MHz below its sweet spot to avoid the unwanted cross talk between $Q_1$ and $Q_3$. The dashed line (the ``operating" point) represents the mean operating frequency of each qubit during the QST experiment: for $Q_0$, it is fixed at the dashed line; for $Q_1-Q_3$, their frequencies are parametrically modulated around the dashed lines to achieve the required coupling $g'_j$ between neighboring qubits.}
\label{fig:Fig1}
\end{figure}

\section{Results}

\subsection{Theory and experimental system}

We first briefly discuss how to tune the coupling strengths in a one-dimensional chain consisting of \emph{N} capacitively coupled qubits, as shown in Fig.~\ref{fig:Fig1}a. The system Hamiltonian can be described by
\begin{eqnarray}\label{Eq1}
H/\hbar=\sum^{N-1}_{j=0}\frac {\omega_j} {2}\sigma^z_j+\sum^{N-1}_{j=1}g_j(\sigma^+_{j-1}+\sigma^-_{j-1})(\sigma^+_{j}+\sigma^-_{j}),
\end{eqnarray}
where  $\sigma^{z,\pm}_j$ are the Pauli operators on the $j$th qubit $Q_j$ with transition frequency $\omega_j$, and $g_j$ is the static coupling strength between qubits $Q_{j-1}$ and $Q_{j}$. Full tunability of the coupling strength can be achieved by our parametrically modulating the qubits; that is, each $Q_j$ with $1\leq j\leq{N-1}$ is biased by an ac magnetic flux to periodically modulate its frequency as
\begin{eqnarray}\label{Eq2}
\omega_j=\omega_{\mathrm{o}j}+\varepsilon_j \sin(\nu_jt+\varphi_j),
\end{eqnarray}
where $\omega_{\mathrm{o}j}$ is the mean operating frequency, and $\varepsilon_j$, $\nu_j$, and $\varphi_j$ are the modulation amplitude, frequency, and phase, respectively. As the first qubit is not modulated, $\omega_0=\omega_{\mathrm{o}0}$. Ignoring the higher-order oscillating terms, when $\Delta_j=\omega_{\mathrm{o}j}-\omega_{\mathrm{o}(j-1)}$ equals to $\nu_j$ ($-\nu_j$) for odd (even) $j$, we get a chain of qubits with the nearest-neighbor resonant $XY$ coupling in the interaction picture. Then the effective Hamiltonian (see Appendix A) is
\begin{eqnarray}\label{eq:H}
H_I/\hbar=\sum^{N-1}_{j=1} g^\prime_j \sigma^+_{j-1}\sigma^-_j+\text{H.c.},
\end{eqnarray}
where the effective coupling strength
\begin{eqnarray}\label{eq:tunable_g}
g^\prime_j= g_j J_1(\alpha_j)\times
\left\{
  \begin{array}{ll}
   e^{i(\varphi_1+\pi/2)}, & \hbox{$j=1$;} \\
    J_0(\alpha_{j-1}) e^{-i(\varphi_j-\pi/2)}, & \hbox{$j$ is even;}\\
   J_0(\alpha_{j-1}) e^{i(\varphi_j+\pi/2)}, & \hbox{$j$ is odd and $\neq$ 1},
      \end{array}
\right.
\end{eqnarray}
with $J_m(\alpha_j)$ being the $m$th Bessel function of the first kind. We can conveniently tune $g^\prime_j$ by changing  $\alpha_j=\varepsilon_j/\nu_j$ of the external modulation.

With a wide range tunability of the coupling strength in hand, we now turn to the demonstration of perfect QST along a chain of qubits~\cite{qst2004}. Initially, we prepare the $i$th qubit $Q_i$ in state $\ket{\psi_i}=\alpha |g\rangle +\beta |e\rangle$,
and all other qubits are in the ground state $|g\rangle$ ($\ket{e}$ represents the excited state). To realize perfect QST, the coupling strengths need to fulfill the relation $g^\prime_j= g^\prime \sqrt{j(N-j)}$~\cite{qst2004}, where $g'$ is a constant. When the system evolves under the Hamiltonian in Eq.~\eqref{eq:H} for a specific time $\tau = \pi/(2g^\prime)$,  perfect QST is achieved, such that qubit $Q_{N-1-i}$ is in the state $\ket{\psi_{N-1-i}}=\ket{\psi_i}$, while all other qubits are back in $|g\rangle$. In our experiment, we demonstrate the case of $N=4$ and $i=0$.

\begin{figure}
\includegraphics{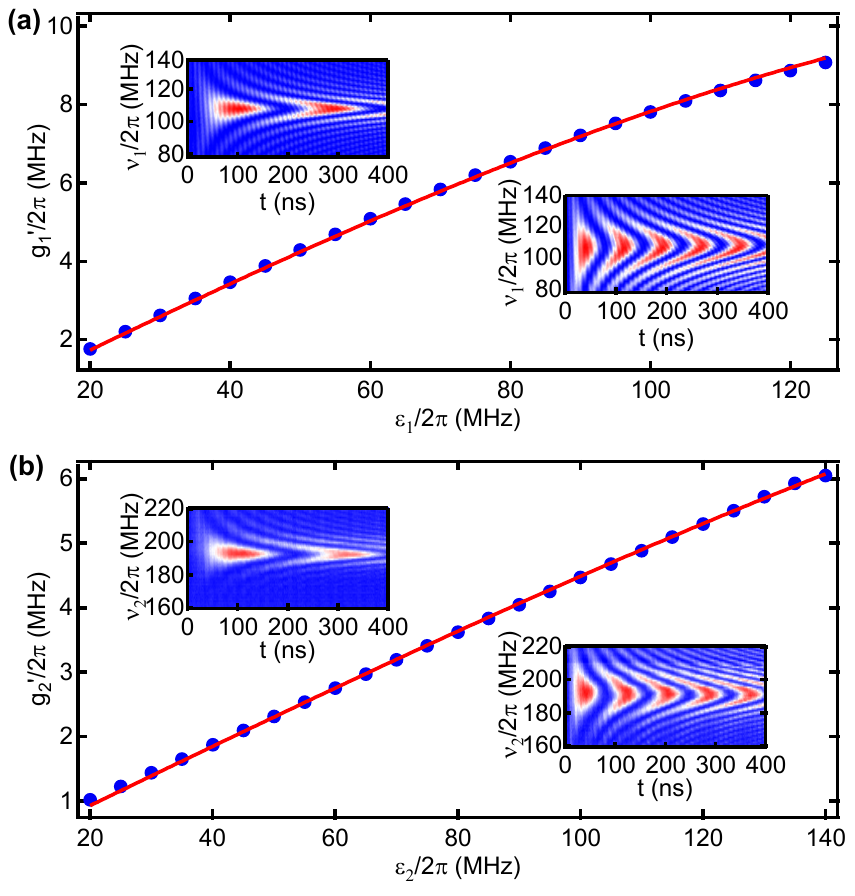}
\caption{ (a) Tunable coupling $g_1'$ between $Q_0$ and $Q_1$ as a function of $\varepsilon_1$ with other qubits being decoupled. $g_1'$ is extracted from the so-called chevron pattern. $Q_0$ is prepared in $\ket{e}$ and biased at a fixed point $\omega_{\mathrm{o0}}$, while $Q_1$, initially in $\ket{g}$, is flux biased to oscillate sinusoidally as $\omega_1=\omega_{\mathrm{o1}}+\varepsilon_1 \sin(\nu_1t+\varphi_1)$. Coherent excitation oscillation between $Q_0$ and $Q_1$ as a function of $\nu_1$ and time $t$ at fixed $\varepsilon_1$ produces a chevron pattern. The full-scale oscillation is achieved when $\nu_1=\Delta_1=\omega_{\mathrm{o1}}-\omega_{\mathrm{o0}}$, and its frequency gives the effective coupling $g_1'$ (dots). Two typical chevron patterns are shown in the inset, with blue and red corresponding to $\ket{g}$ and $\ket{e}$, respectively,  of $Q_1$: the top-left pattern and the bottom-right pattern correspond to $\varepsilon_1/2\pi=30~$MHz and $\varepsilon_1/2\pi=80~$MHz, respectively.  (b) Tunable coupling $g_2'$ between $Q_1$ and $Q_2$ as a function of $\varepsilon_2$ with other qubits being decoupled.  Both $Q_1$ (initially in $\ket{e}$) and $Q_2$ (initially in $\ket{g}$) are parametrically driven as $\omega_1=\omega_{\mathrm{o1}}+\varepsilon_1 \sin(\nu_1t+\varphi_1)$ and $\omega_2=\omega_{\mathrm{o2}}+\varepsilon_2 \sin(\nu_2t+\varphi_2)$ simultaneously. Similarly, at $\nu_2=\Delta_2=\omega_{\mathrm{o2}}-\omega_{\mathrm{o1}}$, the oscillation frequency gives the effective coupling $g_2'$ (dots). The top-left and bottom-right insets show the chevron pattern with $\varepsilon_2/2\pi=50~$MHz and $\varepsilon_2/2\pi=140~$MHz, respectively. The red lines in (a) and (b) are fitted with Eq.~(\ref{eq:tunable_g}). The resulting $g_1/2\pi \approx g_2/2\pi \approx 19$~MHz is slightly larger than the value measured from a static case in which the chevron pattern is measured while  the frequency of one qubit is tuned across the other qubit with no parametric modulation. The deviation is presumably due to imperfect deconvolution of the flux pulse (see Appendix D), so the effective $\varepsilon_1$ and $\varepsilon_2$ are slightly larger than applied.}
\label{fig:Fig2}
\end{figure}

\begin{figure}[t]
\includegraphics{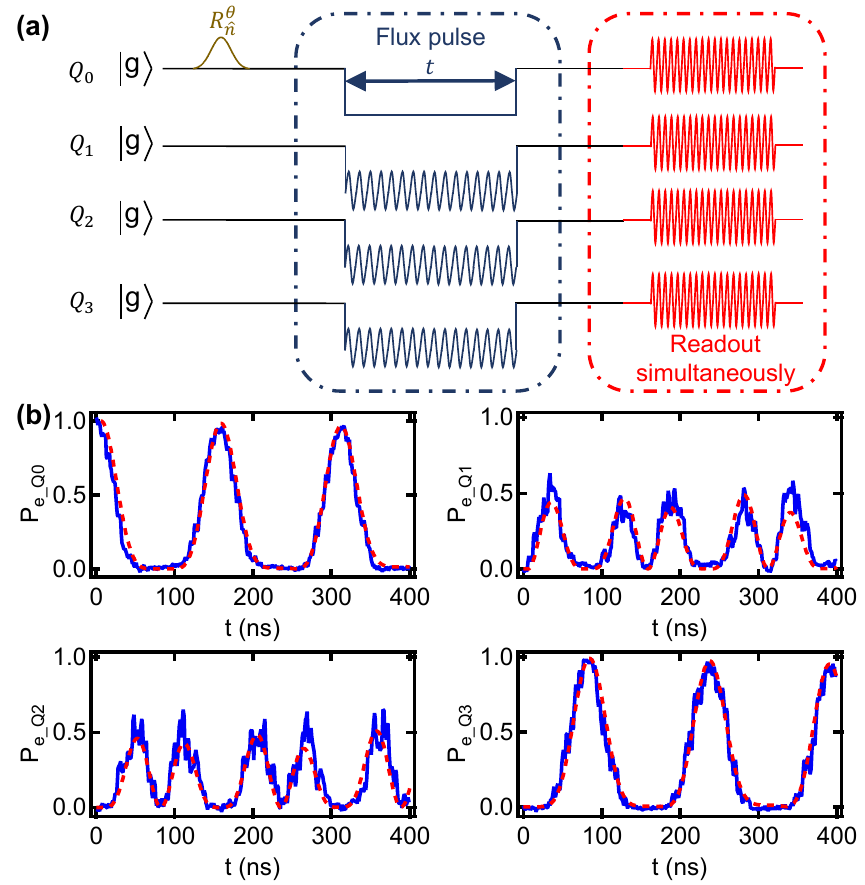}
\caption{ (a) Experimental sequence for QST.  Initially, all four qubits are at their idle points, with only $Q_0$ prepared in an arbitrary state by $R^{\theta}_{\hat{n}}$. Then step pulses are used to change the qubits from their idle spots to the operating points. $Q_1-Q_3$ are modulated to achieve the required coupling ratio for perfect QST. The operation is on for various times $t$, followed by step pulses to return all qubits to their idle points for the final-qubit-state read-out. (b) Time evolution of the qubit excited-state populations. The populations are measured simultaneously at different $t$ for the case with $Q_0$ initially in $\ket{e}$ as an example to calibrate the QST time. The dashed red lines are numerically simulated results with the measured parameters based on the Hamiltonian in Eq.~(\ref{eq:H}). The population evolutions of $Q_1$ and $Q_2$ are noisier because in our implementation $Q_1$ and $Q_2$ are affected by multiple modulations, which lead to more high-order oscillations and cross talk. The slight deviation between simulation and experimental results mainly comes from the high-order terms and imperfect read-out calibration matrix due to the unwanted cross talk between qubits. At $t=84$~ns,  QST from $Q_0$ to $Q_3$ is achieved.}
\label{fig:Fig3}
\end{figure}

\begin{figure}
\includegraphics{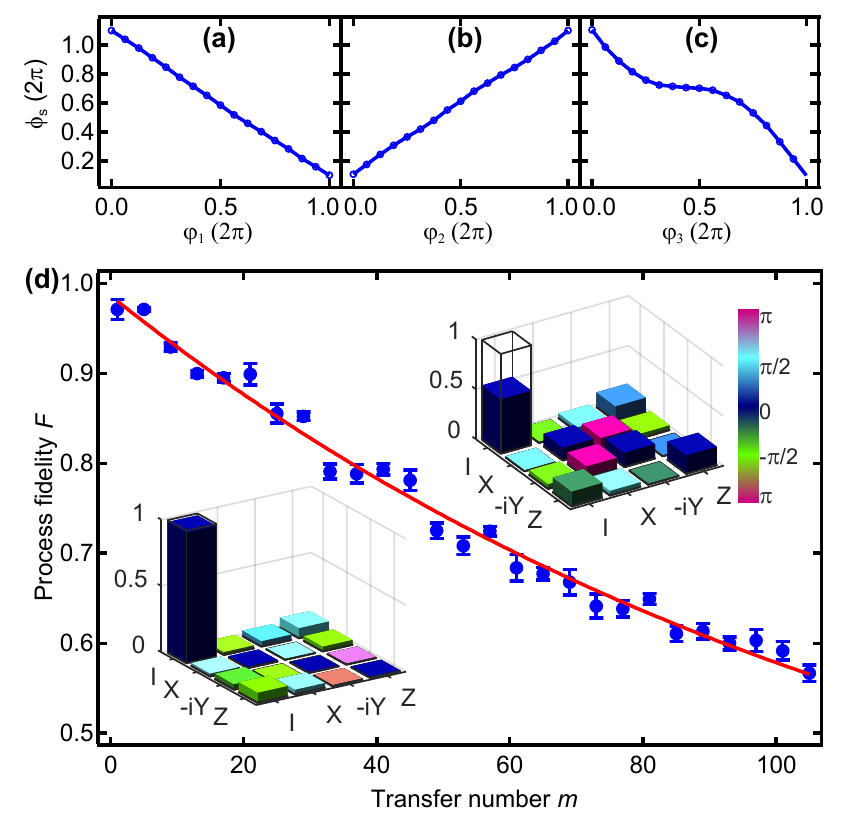}
\caption{ (a-c) Dependence of phase $\phi_s$ of the transferred state on $\varphi_1$, $\varphi_2$, and $\varphi_3$ of the modulation pulses. The dependences of $\phi_s$ on $\varphi_1$ and $\varphi_2$ are linear but with opposite slopes as expected. The dependence of $\phi_s$ on $\varphi_3$ deviates from a linear curve due to an extra phase accumulation. (d) QST process fidelity. Process tomography is used to benchmark the QST performance and is plotted as a function of the number of transfers, which is controlled by our setting the operating times $t=(4n+1)\times 84$~ns, equivalent to performing $1,5,9,\cdots$ transfer processes. The phase of the transferred state on $Q_3$ is deterministic and can be well controlled by adjustment of  the phases in the sinusoidal modulation pulses as shown in  (a)-(c). The process fidelities presented are all based on the transferred states after proper phase adjustments. The parameters $\varepsilon_i$ and $\nu_i~(i=1,2,3)$ are critical for high-fidelity QST and thus are further optimized by the function fminsearch  in MATLAB. Each point is averaged 10 000 times and the error bars corresponding to one standard deviation are obtained from ten repeated experiments. The red curve is a fit based on $F=AP^m+0.25$ with $A=0.737$ and $P=0.992$, demonstrating nearly perfect QST. The bottom-left inset shows the $\chi$ matrix after one QST and the top-right inset shows the $\chi$ matrix after  105 QSTs. The bar height and color correspond to the amplitude and phase of the $\chi$-matrix element, respectively.}
\label{fig:Fig4}
\end{figure}

%\section{Experimental results}
Our experiment is implemented with a superconducting circuit~\cite{Clarke2008,You2011,Devoret2013,GU2017}, where five cross-shaped transmon qubits (Xmons, $Q_0-Q_4$)~\cite{BarendsPRL2013,Barends2014,Kelly2015} are arranged in a linear array with nearly identical nearest-neighbor coupling strengths $g/2\pi\approx 17$~MHz, as shown in Fig.~\ref{fig:Fig1}b. Each qubit has independent $XY$ and $Z$ control and is coupled to a separate $\lambda/4$ resonator for simultaneous and individual read-out. The qubits have averaged $T_1\approx 22~\mu$s and $T^*_2\approx 19~\mu$s at the flux sweet spots. A Josephson parametric amplifier (JPA)~\cite{Hatridge,Roy2015,Kamal,Murch} with a gain of more than 20~dB and a bandwidth of about 260~MHz is used for high-fidelity single-shot measurements of the qubits. Meanwhile, we use a calibration matrix to reconstruct the read-out results for a better indication of the qubit state. The experimental setup, read-out properties, and device parameters are presented in Appendixes B and C.

\subsection{Coupling tunability through parametric modulations}

As a demonstration, we use the first four qubits ($Q_0-Q_3$) to realize QST, while biasing $Q_4$ at a low frequency (less than 4~GHz); $Q_4$ is nearly completely decoupled from the first four qubits. Synchronization and phase stability of both the $XY$ control and the $Z$ control  are critical for high-fidelity QST. Therefore, in our experiment, we use two synchronized four-channel arbitrary-waveform generators (AWGs) to fully manipulate the four qubits. Figure~\ref{fig:Fig1}c shows the biasing and operating regime of the four qubits. Both state preparation and final measurement of each qubit are performed at or near the maximum-frequency spot (the idle point at $\omega_{\mathrm{s}j}$ with $j=$ 0, 1, 2, 3) with the best coherence times. During the QST experiment, all qubits are pulsed to the ``operating" points, $\omega_{\mathrm{o}j}$. While $Q_0$ stays fixed, the other three qubits are parametrically driven to oscillate sinusoidally around their operating points to achieve the required coupling $g'_j$ between neighboring qubits.

We first demonstrate the tunability of $g'_j$ by parametric modulations of the transition frequencies of the qubits. There are two scenarios of tuning $g'_j$: (a) one qubit is modulated while the other one remains at a fixed frequency; (b) both qubits are modulated simultaneously. The former case of parametric modulation has been used to create entangling gates between two transmon qubits~\cite{Caldwell2017,Reagor2018,Didier2018}. However, to our knowledge the latter case with simultaneous parametric modulations on two qubits has not yet been demonstrated experimentally. Figures~\ref{fig:Fig2}a and \ref{fig:Fig2}b show the experimental results of the two scenarios, demonstrating smooth and full control of $g'_j$ as a function of $\varepsilon_j$.

\subsection{Calibration and performance of perfect QST}

Once proper couplings $g'_j$  between neighboring qubits in the chain are achieved by appropriate qubit frequency modulations, perfect QST can be calibrated by our measuring the population of each qubit as a function of time. Figure~\ref{fig:Fig3}a shows the experimental sequence. Initially, all four qubits are at their idle points and in the ground state except for $Q_0$, which is prepared in an arbitrary state. Then step pulses are used to change the qubits from their idle spots to the operating points. $Q_1-Q_3$ are frequency modulated with the calibrated $\varepsilon'$ and $\nu'$ such that $g_1':g_2':g_3'=\sqrt{3}:2:\sqrt{3}$ as required for perfect QST~\cite{qst2004}. The operation is for various times $t$, followed by step pulses to return all qubits to their idle points for the final-state read-out. Figure~\ref{fig:Fig3}b shows the measured qubit populations as a function of time with $Q_0$ prepared in $\ket{e}$ as an example. As expected, the population of $Q_0$ first spreads to $Q_1$, $Q_2$, and $Q_3$. At $t=84$~ns, the population is transferred to $Q_3$, while  $Q_0$, $Q_1$, and $Q_2$ go back to the ground states, realizing fast and nearly perfect QST. As $t$ becomes longer, the reverse process occurs.  At $t=168$~ns, the population of $Q_3$ is transferred back to $Q_0$. This transfer can keep going back and forth as $t$ increases. As shown below, this property allows us to better calibrate the QST's process fidelity with repeated transfers. This method  of calibration can eliminate the detrimental effect in the ``round trip'' of the qubit frequency between the idle and operating points, and focuses only on the accumulation of errors during the transfer.

The phase of the transferred state $\phi_s$ depends on the three phases $\varphi_1$, $\varphi_2$, and $\varphi_3$ of the parametric modulations as shown in Figs.~\ref{fig:Fig4}a-c, where the $\varphi$ are the phases in the corresponding sinusoidal flux drives. $\phi_s$ is a linear function of $\varphi_1$ and $\varphi_2$ as expected from Eq.~\eqref{eq:tunable_g}. However, the dependence of $\phi_s$ on $\varphi_3$ deviates from the expected linear curve. This is because $\phi_s$ also includes an extra phase accumulation during $Q_3$'s frequency modulation when $\varphi_3$ changes. Nevertheless, $\phi_s$ can be fully controlled by $\varphi_1$, $\varphi_2$, and $\varphi_3$ individually.

Quantum process tomography~\cite{Nielsen,Chow2009} is used to benchmark the QST performance and the fidelity is defined as the overlap between $\chi_{\mathrm{M}}$ and $\chi_{\mathrm{ideal}}$ ($\chi_{\mathrm{ideal}}$ is for perfect QST) $F=\mathrm{tr}(\chi_{\mathrm{M}}\chi_{\mathrm{ideal}})$, where $\chi_{\mathrm{M}}$ is the derived $4\times4$ process matrix for the experimental operations on four different initial states, $\{\left|g\right\rangle , \left|e\right\rangle, (\left|g\right\rangle +\left|e\right\rangle)/\sqrt{2}, (\left|g\right\rangle -i\left|e\right\rangle)/\sqrt{2}\}$. Figure~\ref{fig:Fig4}d shows the measured process fidelity as a function of the number of transfers. The number of transfers is controlled by our properly setting the operating time. A fit (red curve) based on $F=AP^m+0.25$ gives $A=0.737$ and $P=0.992$ without relying on perfect state preparation and perfect measurement, demonstrating nearly perfect QST. This high-fidelity process, dominantly limited by qubit decoherence, is possible mainly because the fast-QST approach requires only a single step, which minimizes the qubit decoherence effect and is robust to noise and accumulation of experimental errors. The pure dephasing time of the transferred state in the QST is much longer than the average dephasing time of individual qubits at the operating points (see Appendix C), and this implies the collective dynamical process provides additional coherence protection. We leave a more detailed study of this protection for future work.

\section{Conclusion}

In conclusion, we demonstrate  fast and high-fidelity QST with four superconducting qubits arranged in an array with nearest-neighbor coupling. This transfer relies on full control of the effective couplings between neighboring qubits by our parametrically modulating the qubits. This tunable technique can be extended to a much-larger system, and importantly the transfer can be achieved in a single step, and is therefore robust to noise and accumulation of experimental errors. The coupling tunability can be realized \emph{in situ} without the circuit complexity being increased, and therefore provides a powerful and desirable tool for future quantum information processing.

Our experiment can be easily extended to achieve entanglement distribution between remote qubits in a chain~\cite{Chapman2016}. Our technique can also be directly generalized to study topologically protected QST from one end to another~\cite{Mei2017}, which requires different configurations of the coupling strengths. In addition, our scheme can be generalized to a two-dimensional lattice, where the tunable amplitudes and phases of the coupling strengths allow  quantum simulation of lattice models~\cite{Georgescu2014,Goldman2014}.

\section*{Acknowledgments}
This work was supported by the National Key R \& D Program of China (Grants No. 2017YFA0304303 and No. 2016YFA0301803) and the National Natural Science Foundation of China (Grants No. 11474177, No. 11874156, and No. 61771278). L. S. thanks R. Vijay and his group for help with the parametric amplifier measurements.

\appendix
\section{Tunable interaction}
Here we present details of how to tune the coupling strength of two adjacent qubits in a chain of $N$ coupled qubits. A wide-range coupling tunability can be achieved through parametric modulations of the qubits $\omega_j=\omega_{\mathrm{o}j}+\varepsilon_j \sin(\nu_jt+\varphi_j)$ with $j=1,...,N-1$. We define a rotating frame through $U=U_1 \times U_2$ with
\begin{eqnarray}\label{Eq1}
U_1&=&\exp\left[-i\left(\frac {\omega_0} {2}\sigma^z_0+\sum^{N-1}_{j=1}\frac {\omega_{oj}} {2}\sigma^z_j\right)t\right]
\end{eqnarray}
and
\begin{eqnarray}\label{Eq2}
U_2&=&\exp\left[i\sum^{N-1}_{j=1}\sigma^z_j\frac {\alpha_j} {2}\cos(\nu_jt+\varphi_j)\right],
\end{eqnarray}
where $\alpha_j=\varepsilon_j/\nu_j$. In this rotating frame, the transformed Hamiltonian is
\begin{eqnarray}\label{Eq3}
H_I &=& U^\dag HU+i\frac {dU^\dag} {dt}U \notag \\
%&=& U^\dag\left[\sum^{N-1}_{j=1}g_j(\sigma^+_{j-1}+\sigma^-_{j-1})(\sigma^+_{j}+\sigma^-_{j})\right]U \notag \\
&=& g_1\sigma^+_0\sigma^-_1 e^{-i\Delta_1 t} \exp\left[i\alpha_1 \cos(\nu_1t+\varphi_1)\right] \notag \\
    & +& \sum^{N-1}_{j=2}g_j \sigma^+_{j-1}\sigma^-_j \exp\left[-i\alpha_{j-1} \cos(\nu_{j-1}t+\varphi_{j-1})\right] \notag\\
&& \times \exp\left[i\alpha_{j} \cos(\nu_{j}t+\varphi_{j})-i\Delta_j t\right] +\text{H.c.}.
\end{eqnarray}
When $\Delta_j=\omega_{\mathrm{o}j}-\omega_{\mathrm{o}(j-1)}$ equals $\nu_j$ ($-\nu_j$) for odd (even) $j$, if we use the Jacobi-Anger identity \begin{eqnarray}\label{Eq4}
\exp[i\alpha \cos(\nu t+\varphi)]=\sum^{\infty}_{m=-\infty}i^m J_m(\alpha)\exp[im(\nu t+\varphi)]\notag
\end{eqnarray}
and apply the rotating-wave approximation by ignoring the high-order oscillating terms, the effective tunable resonant qubit-qubit interaction becomes Eq.~(3).

\section{Experimental Device}
%\subsection{Fabrication}
Fabrication of the experimental device includes the following six main steps: (1) A 100-nm aluminum film is deposited directly onto a 2-in. $c$-plane sapphire wafer in Plassys MEB 550S without any precleaning treatment. (2) Electron-beam lithography followed by evaporation of gold is used to create alignment marks for the subsequent lithography steps. (3) Photolithography followed by inductively-coupled-plasma etching is used to define transmission line, read-out resonators, control lines, and large pads of the Xmon qubits. (4) Josephson junctions are fabricated by electron-beam lithography and double-angle evaporation of aluminum in Plassys MEB 550S. (5) The wafer is then diced into $7\times7$ mm$^2$ chips. (6) The selected chip is wire bonded in an aluminum box without a printed circuit board for final packaging before measurement. We apply as many on-chip bonding-wire crossovers as  possible to reduce the impact of the parasitic modes.

\begin{figure*}[hbt]
{\centering\includegraphics[width=0.74\textwidth]{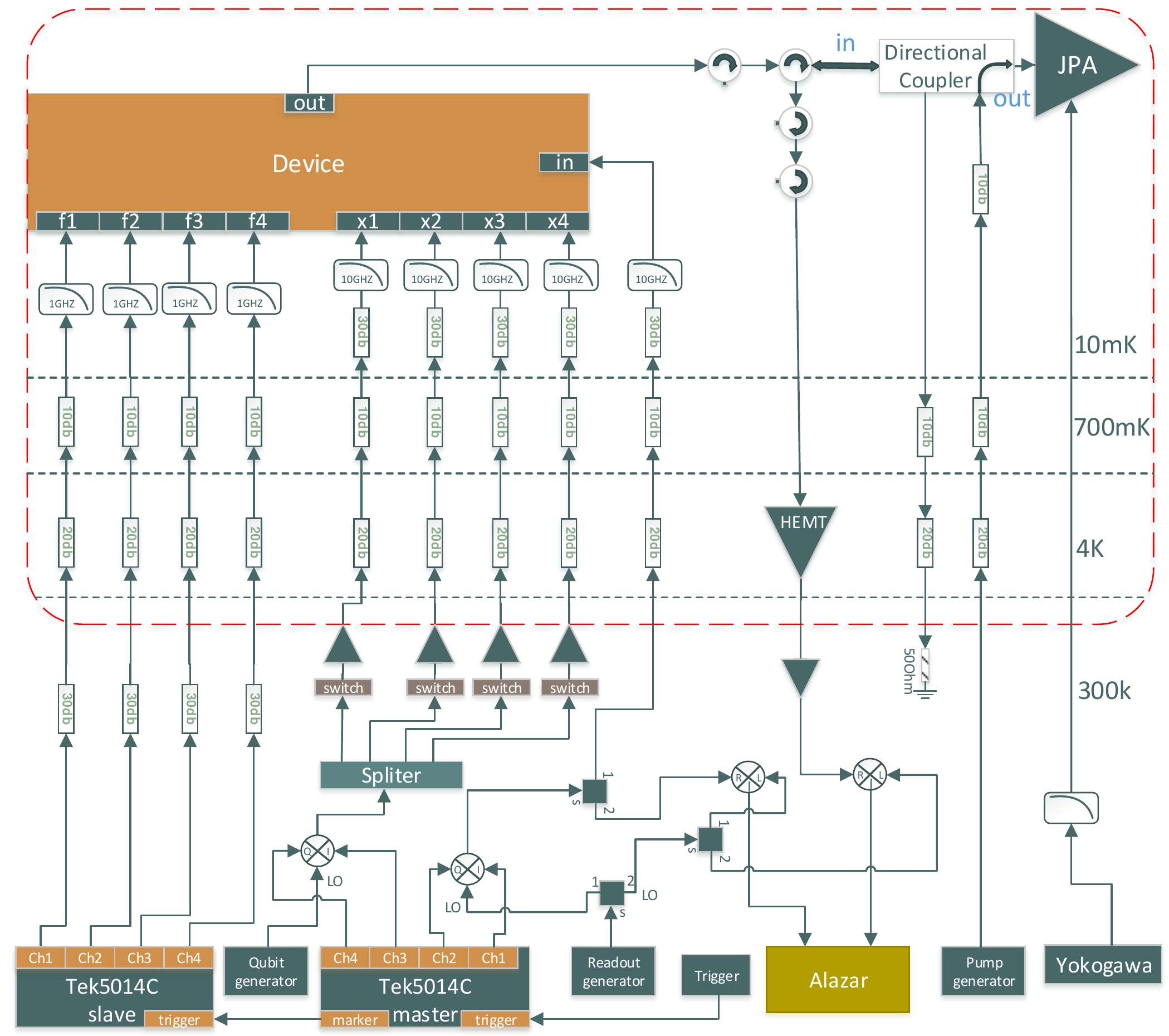}}
\caption{ Details of wiring and circuit components. The experimental device consists of five cross-shaped transmon qubits (Xmons, $Q_0-Q_4$)~\cite{BarendsPRL2013,Barends2014,Kelly2015} arranged in a linear array with nearly identical nearest-neighbor coupling strengths. Each qubit has independent $XY$ and $Z$ controls, which are properly attenuated and low-pass filtered. A common transmission line is coupled to separate $\lambda/4$ resonators for individual read-outs of the qubits. Two four-channel AWGs are used to fully manipulate the four qubits ($Q_0-Q_3$) to realize the QST. The fifth qubit, $Q_4$, is biased with a dc source at a low frequency (less than 4~GHz) and is nearly completely decoupled from the first four qubits. The master AWG provides two pairs of sideband modulations at different frequencies, in combination with a qubit generator and a read-out generator as local oscillators (LOs), allowing $XY$ control and read-out of the qubits, respectively. The $XY$ control signal is divided by a four-way power divider, and the outputs are connected to the respective qubit $XY$ control lines through separate rf switches. These switches provide selective control of individual qubits. The slave  AWG, triggered by the master AWG, is use to achieve $Z$ control of the qubits through individual flux-bias lines. A JPA at 10~mK with a gain of more than 20~dB and a bandwidth of about 260~MHz is used as the first stage of amplification, allowing high-fidelity single-shot measurements of the qubits. A high-electron-mobility-transistor (HEMT) amplifier at 4 K and an amplifier at room temperature are also used before the down-conversion of the read-out signal to the applied sideband frequencies with the same read-out generator as the LO. Part of the read-out signal does not go through the dilution refrigerator and is used as a reference to lock the phase of the returning read-out signal from the device for greater measurement stability.}
\label{fig:experimentsetup}
\end{figure*}

\begin{figure}[b]
\centering
\includegraphics[scale=0.9]{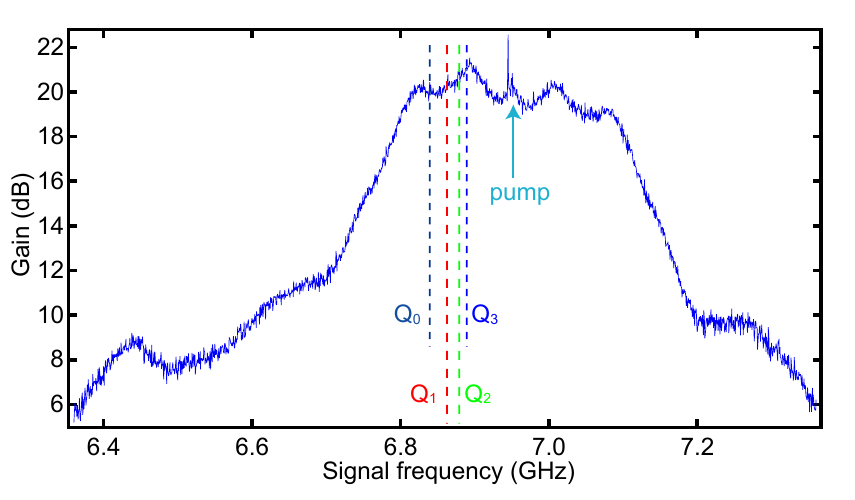}
\caption{Gain profile of the JPA. The JPA is singly  pumped at a frequency of 6.951~GHz. The dashed lines represent the read-out frequencies for $Q_0$-$Q_3$ respectively. The maximum gain is greater than 20~dB and the bandwidth is about 260~ MHz.}
\label{fig:Amplifier gain}
\end{figure}

%\subsection{Diagram of experimental setup}
Our sample is measured in a dilution refrigerator with a base temperature of about 10~mK. Details of our measurement circuitry are shown in Fig.~\ref{fig:experimentsetup}. For full manipulation of four qubits, we use two four-channel  AWGs. One AWG provides two pairs of sideband modulations for $XY$ control and read-out of the qubits, respectively. The second AWG, synchronized with the first one, is used to realize individual $Z$ control of the qubits. The $XY$ control signals are generated from a single microwave generator modulated with different sideband frequencies. This method of control guarantees stable phase differences among the four qubits during the  QST experiment. The read-out signals for individual qubits are realized in a similar way to make sure that the phases of the final demodulated readout signals are fixed for each run. A JPA~\cite{Hatridge,Roy2015,Kamal,Murch} at 10~mK with a gain of more than 20~dB and a bandwidth of about 260~MHz is used as the first stage of amplification, allowing high-fidelity single-shot measurements of the qubits. Its gain profile is shown in Fig.~\ref{fig:Amplifier gain}. We do not apply any specific impedance engineering in the JPA circuit other than 50-$\Omega$ impedance matching, and the high bandwidth is unintentional.

Read-out-resonator frequencies, qubit frequencies, qubit coherence times, coupling strengths, and read-out-resonator decay rates are all presented in Table~\ref{Table:parameters}. The readout frequencies of the four qubits span a range of about 60~MHz, well within the bandwidth of the JPA. In the current device, the dispersive shifts $\chi_{\mathrm{qr}}$ (between each qubit and its read-out resonator) and the read-out-resonator decay rates $\kappa_{\mathrm{r}}$ are not matched for the best signal-to-noise ratio.

\begin{table*}[bht]
\caption{ Device parameters.  The parameters associated with qubit $Q_4$ are not presented since it is not involved in the QST experiment.}
%\begin{tabular*}{1\textwidth}{@{\extracolsep{\fill}}@{\extracolsep{\fill}}|c|c|c|c|c|c|c|c|c}
\begin{tabular}{cp{1.2cm}<{\centering}p{1.2cm}<{\centering}p{1.2cm}<{\centering}p{1.2cm}<{\centering}p{1.2cm}<{\centering}p{1.2cm}<{\centering}p{1.2cm}<{\centering}p{1.2cm}<{\centering}}
%\hline
%1&2&3&4&5&6&7&8&9 \tabularnewline
&&&&&&&& \tabularnewline
\hline
\hline
Parameter &\multicolumn{2}{c}{$Q_0$}  &\multicolumn{2}{c}{$Q_1$} &\multicolumn{2}{c}{$Q_2$} &\multicolumn{2}{c}{$Q_3$} \tabularnewline
\hline
Read-out frequency (GHz) &\multicolumn{2}{c}{$6.8389$} &\multicolumn{2}{c}{$6.8636$} &\multicolumn{2}{c}{$6.8794$} &\multicolumn{2}{c}{$6.9014$} \tabularnewline
%\hline
Qubit frequency (GHz) (sweet spot) &\multicolumn{2}{c}{$4.8354$} &\multicolumn{2}{c}{$5.1802$} &\multicolumn{2}{c}{$4.9169$} &\multicolumn{2}{c}{$5.1916$} \tabularnewline
%\hline
$T_1$ ($\mu$s) (sweet spot) &\multicolumn{2}{c}{$22.2$} &\multicolumn{2}{c}{$18.5$} &\multicolumn{2}{c}{$25.1$} &\multicolumn{2}{c}{$23.4$} \tabularnewline
%\hline
$T^*_2$ ($\mu$s) (sweet spot) &\multicolumn{2}{c}{$23.3$} &\multicolumn{2}{c}{$26.5$} &\multicolumn{2}{c}{$17.3$} &\multicolumn{2}{c}{$10.3$} \tabularnewline
%\hline
$T_{\mathrm{2E}}$ ($\mu$s) (sweet spot) &\multicolumn{2}{c}{$24.0$} &\multicolumn{2}{c}{$41.1$} &\multicolumn{2}{c}{$29.3$} &\multicolumn{2}{c}{$32.2$} \tabularnewline
%\hline
$T_1$ ($\mu$s) (operating point) &\multicolumn{2}{c}{$17.5$} &\multicolumn{2}{c}{$21.1$} &\multicolumn{2}{c}{$19.8$} &\multicolumn{2}{c}{$18.0$} \tabularnewline
%\hline
$T^*_2$ ($\mu$s) (operating point) &\multicolumn{2}{c}{$6.1$} &\multicolumn{2}{c}{$4.3$} &\multicolumn{2}{c}{$4.8$} &\multicolumn{2}{{c}}{$3.3$} \tabularnewline
%\hline
Neighboring-qubit coupling strength ${\rm g}_{j}/2\pi$ (MHz) & &\multicolumn{2}{c}{$16.68$} &\multicolumn{2}{c}{$17.50$} &\multicolumn{2}{c}{$17.52$} & \tabularnewline
Qubit-read-out-resonator dispersive shift $\rm\chi_{qr}/2\pi$ (MHz) &\multicolumn{2}{c}{$0.17$} &\multicolumn{2}{c}{$0.26$} &\multicolumn{2}{c}{$0.2$} &\multicolumn{2}{{c}}{$0.2$} \tabularnewline
Read-out-resonator decay rate $\rm\kappa_r/2\pi$ (MHz) &\multicolumn{2}{c}{$0.88$} &\multicolumn{2}{c}{$1.06$} &\multicolumn{2}{c}{$1.23$} &\multicolumn{2}{{c}}{$0.88$} \tabularnewline
\hline
\hline
\end{tabular} \vspace{-6pt}
\label{Table:parameters}
\end{table*}

\section{Qubit Read-out Properties and Process Tomography}

\begin{table}[t]
\caption{ The read-out fidelities for each qubit. The fidelities are based on the histograms presented in Fig.~\ref{fig:ReadoutHistogram}. $F_{\rm{g}}$ is measured for an initial thermal steady state. $F_{\rm{e}}$ corresponds to an initial thermal steady state followed by the corresponding $\pi$ rotation.}
\begin{center}
\begin{tabular}{p{1.5cm}<{\centering}p{1.5cm}<{\centering}p{1.5cm}<{\centering}p{1.5cm}<{\centering}p{1.5cm}<{\centering}}
\hline
\hline
\centering
       & $Q_0$  & $Q_1   $ & $Q_2$ & $Q_3$ \\
\hline
$F_{{\rm g}j}$  & 0.963  & 0.951 & 0.942 & 0.939 \\
$F_{{\rm e}j}$   & 0.898  & 0.869 & 0.869 & 0.858   \\
\hline
\hline
\end{tabular} \vspace{8pt}\\
\label{Table:ReadoutFidelity}
\end{center}
\end{table}

\begin{figure}[t]
\centering
\includegraphics[scale=1.01]{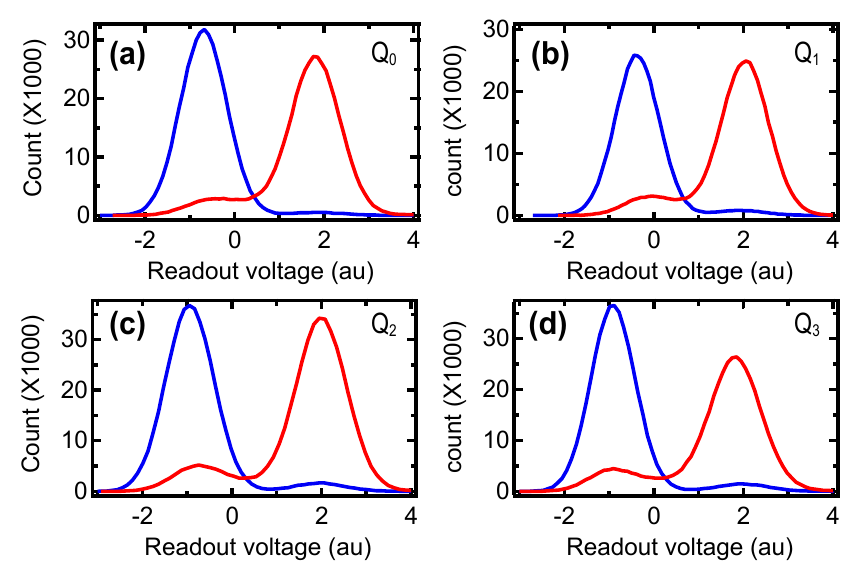}
\caption{Read-out histograms. (a)-(d) Histograms for $Q_0$-$Q_3$ respectively. Each histogram is measured separately with a total count of 90 000, while the other qubits are in their thermal steady states. Blue curves are for an initial thermal steady state and red curves are for an initial thermal steady state followed by the corresponding $\pi$ rotation.}
\label{fig:ReadoutHistogram}
\end{figure}

With the help of the JPA, all qubits at the idle points can be read out individually with high fidelities. Figure~\ref{fig:ReadoutHistogram} shows the read-out histograms of each qubit when all other qubits are in their thermal steady states. Because of the mismatch of the dispersive shift and the read-out-resonator decay rate, the two Gaussians in the histograms corresponding to the ground state $\ket{g}$ and the excited state $\ket{e}$ are not perfectly separated. The read-out fidelities $F_{\rm{g}}$ and $F_{\rm{e}}$ for each qubit are given in Table~\ref{Table:ReadoutFidelity}. Here $F_{\rm{g}}$ is measured for an initial thermal steady state, while $F_{\rm{e}}$ corresponds to an initial thermal steady state followed by the corresponding $\pi$ rotation. The infidelity of $F_{\rm{g}}$ mainly comes from the thermal population of the qubit (about 0.02 on average from an independent measurement) and the Gaussian tail after thresholding. The lower value of $F_e$ is dominantly due to an extra decay during the measurement time, 2~$\mu$s in our experiment. To overcome these imperfections, we use a calibration matrix to reconstruct the read-out results based on Bayes's rule.

For the $j$th qubit, we have the calibration matrix
\begin{center}\label{S1}
${F}_{{\rm Q}j}$=
$\left(\begin{tabular}{cc}
${F}_{{\rm g}j}$ & $1-{F}_{{\rm e}j}$\\
$1-{F}_{{\rm g}j}$ & ${F}_{{\rm e}j}$\\
\end{tabular}\right)$.
\vspace{8pt}%\end{eqnarray}
\end{center}
The final state population of the qubit as a column vector $P_{{\rm f}j}$ can be reconstructed from the measured population $P_{{\rm m}j}$ on the basis of the inverse of the calibration matrix:
\begin{center}
%\begin{eqnarray}
%\begin{equation}\label{S1}
$P_{{\rm f}j}=F^{-1}_{{\rm Q}j}\cdot P_{{\rm m}j}.$\\
%\end{eqnarray}
%\end{equation}
\end{center}

Because of residual $ZZ$ coupling (which depends on the detuning) between qubits, the histograms of each qubit are slightly shifted when the other qubits are not in their thermal steady states. In this case, more-thorough calibrations are needed to get better calibration matrices. We attribute the imperfect read-out calibration matrices to the deviation between experimental and simulation results [Fig.~3(b)]. However, after a perfect QST [Fig.~4(d)], the tomography measurement of $Q_3$ does not have this issue since the other three qubits have all returned to their initial states.

Process tomography is realized by our preparing four linear independent initial states $\{\left|g\right\rangle, \left|e\right\rangle, (\left|g\right\rangle +\left|e\right\rangle )/\sqrt{2}, (\left|g\right\rangle -i\left|e\right\rangle )/\sqrt{2}\}$ on $Q_0$, and performing the corresponding final-state tomography of $Q_3$ after the QST. On the basis of these processes, we derive the $4\times4$ process matrix $\chi_{\mathrm{M}}$~\cite{Nielsen,Chow2009}. The fidelity is defined as the overlap between $\chi_{\mathrm{M}}$ and $\chi_{\mathrm{ideal}}$ ($\chi_{\mathrm{ideal}}$ is for perfect QST), $F=\mathrm{tr}(\chi_{\mathrm{M}}\chi_{\mathrm{ideal}})$.

\begin{figure}[t]
\centering
\includegraphics{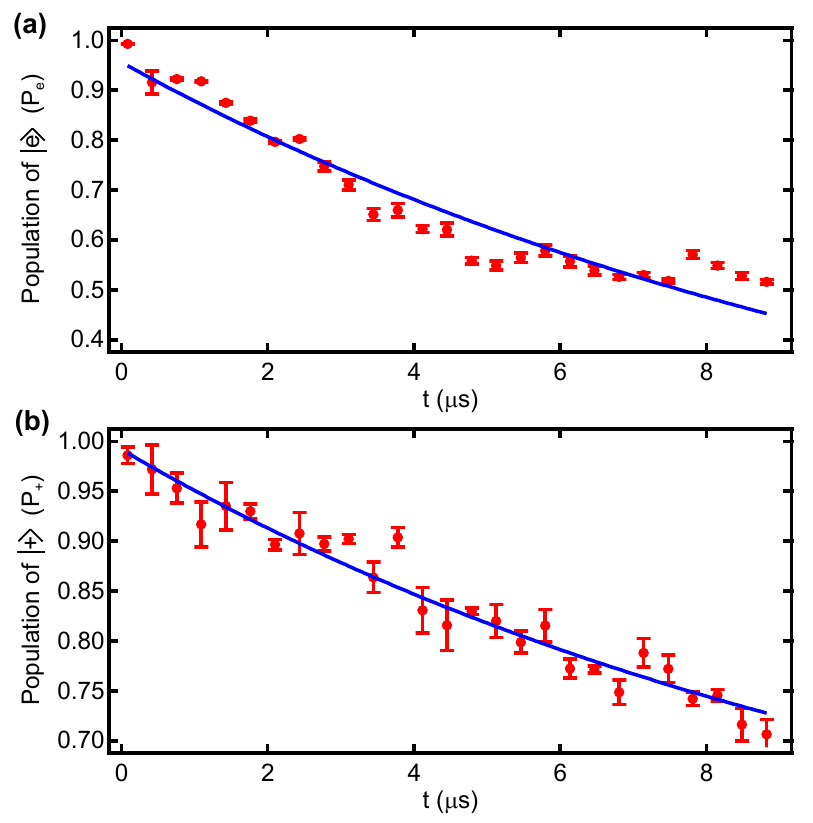}
\caption{(a) and (b) The longitudinal and transverse relaxation of the transferred state with an initial state of $\ket{e}$ and $(\ket{g}+\ket{e})/\sqrt{2}$ $(\ket{+}~\mathrm{state})$, respectively. These data are part of the data presented in Fig.~4(d). Dots are experimental data. Lines are exponential fits, giving the decay times of the transferred state $T_1=11.8~\mu$s and $T^*_2=11.5~\mu$s, respectively.}
\label{fig:T1T2_QST}
\end{figure}

\begin{figure}[b]
\centering
\includegraphics{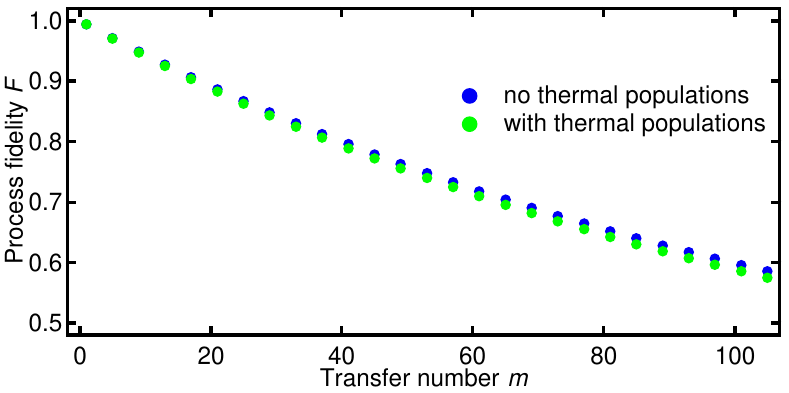}
\caption{Simulated process fidelity $F$ with and without qubit thermal populations (0.02 on average) as a function of the transfer number $m$. Dots are simulated data. Similarly to Fig.~4(d), fits with $F=AP^m+0.25$ give a fidelity difference of only 0.03\% (0.9920 vs 0.9923). In the simulation, we choose $T^*_2=11.5~\mu$s for all qubits and the measured $T_1$ for each qubit at the operating point as in Table~\ref{Table:parameters}.}
\label{fig:Simiulation_thermal}
\end{figure}

The QST in our demonstration is a dynamical process in which all qubits including the middle ones in the chain, participate, as shown in Fig.~3(b). Complicated and entangled states among the qubits are actually involved in the QST process. Therefore, even for an initial $\ket{e}$ state, the decay time for the transferred excitation should be set by not only the energy-relaxation process  but also the dephasing process of the qubits. The decay times of the transferred state for an initial $\ket{e}$ state or $(\ket{g}+\ket{e})/\sqrt{2}$ superposition state should not be very different. Figure~\ref{fig:T1T2_QST} shows the longitudinal and transverse relaxation of the transferred state with an initial state of $\ket{e}$ and $(\ket{g}+\ket{e})/\sqrt{2}$, respectively. These data are part of the data presented in Fig.~4(d). The decay times of the transferred state in both cases are about the same and are much longer than the average dephasing time of the four qubits when they are stationary and at their operating points. This implies the collective dynamical process provides additional coherence protection. A more-detailed study of this observation is beyond the scope of the current work, and we leave it for future studies.

Figure~\ref{fig:Simiulation_thermal} shows the simulated process fidelity with and without qubit thermal populations (0.02 on average) as a function of the transfer number, similar to Fig.~4b of the main text. As shown in Fig.~\ref{fig:Simiulation_thermal}, the qubit thermal populations introduce only negligible effect (a fidelity difference of only 0.03\%) on the measured process fidelity. The measured process fidelity is still dominantly limited by the qubit decoherence.

\begin{figure}[t]
\centering
\includegraphics[scale=1]{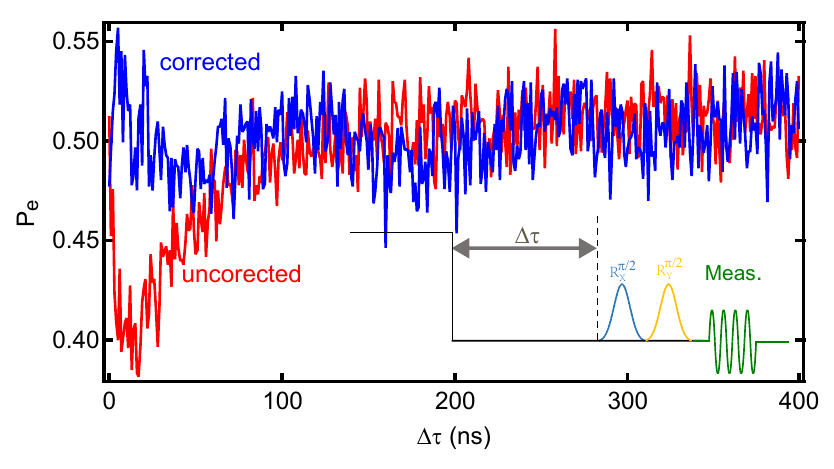}
\caption{ Qubit response to fast flux bias. The red curve corresponds to the uncorrected response. The blue curve corresponds to the corrected response after deconvolution. The inset shows the pulse sequence of the experiment.}
\label{fig:Z_ripple}
\end{figure}

\section{Cross talk and Deconvolution of flux-bias line}
Because of the ground-plane return currents, there is inevitable cross talk (the maximum in our device is about 10\%) between flux-bias lines and qubits. That is, changing the bias on any single qubit's flux line actually changes all of the qubit frequencies. However, this cross talk can be corrected by orthogonalization of the flux-bias lines~\cite{Reedthesis}. For each qubit, we measure its frequency change as a function of the applied voltage on each flux-bias line. For small voltages, the frequency dependence is approximately linear. Therefore, the ratios among the slopes represent the relative coupling strengths between the flux-bias lines and the specific qubit. By combing the results from all four qubits, we can get the qubit frequency response matrix $M_z$. The inverse of $M_z$ gives the orthogonalization matrix $\widetilde{M}_z$, which allows independent control of only the desired qubit without change of  the other qubit frequencies:
\begin{center}
$\widetilde{M}_z$=$M_z^{-1}$=
$\left(\begin{tabular}{cccc}

0.9934 & 0.0822 & 0.021  & 0.0158 \\
-0.0714 & 0.9843  & 0.0595  & 0.0361  \\
-0.0222 & -0.1278  & 0.9888 & 0.074  \\
-0.0087 & -0.057  & -0.0414  & 0.9447  \\

\end{tabular}\right).$%\end{table}
\end{center}

To achieve  high-fidelity QST, flux-bias lines on fast timescales also have to be carefully calibrated to compensate for the finite rise time and ringing of the flux-control pulses. These flux-bias imperfections, seen by the qubit, mainly come from the control circuit, including the AWG to generate those control pulses and wiring outside and inside the refrigerator. We use the deconvolution method to correct the unwanted response in the control system on the basis of  the measured response function of the control circuit~\cite{Johnsonthesis}. The performance of this correction can be verified by measurement of the qubit's response to a step pulse on the flux bias based on the sequence shown in the inset in Fig.~\ref{fig:Z_ripple}~\cite{Kellythesis}. The results shown in Fig.~\ref{fig:Z_ripple} demonstrate an improvement after correction.

\vbox{}

%\noindent \textbf{\large{}{}{}{}Methods}{\large \par}


\begin{thebibliography}{55}%
\makeatletter
\providecommand \@ifxundefined [1]{%
 \@ifx{#1\undefined}
}%
\providecommand \@ifnum [1]{%
 \ifnum #1\expandafter \@firstoftwo
 \else \expandafter \@secondoftwo
 \fi
}%
\providecommand \@ifx [1]{%
 \ifx #1\expandafter \@firstoftwo
 \else \expandafter \@secondoftwo
 \fi
}%
\providecommand \natexlab [1]{#1}%
\providecommand \enquote  [1]{``#1''}%
\providecommand \bibnamefont  [1]{#1}%
\providecommand \bibfnamefont [1]{#1}%
\providecommand \citenamefont [1]{#1}%
\providecommand \href@noop [0]{\@secondoftwo}%
\providecommand \href [0]{\begingroup \@sanitize@url \@href}%
\providecommand \@href[1]{\@@startlink{#1}\@@href}%
\providecommand \@@href[1]{\endgroup#1\@@endlink}%
\providecommand \@sanitize@url [0]{\catcode `\\12\catcode `\$12\catcode
  `\&12\catcode `\#12\catcode `\^12\catcode `\_12\catcode `\%12\relax}%
\providecommand \@@startlink[1]{}%
\providecommand \@@endlink[0]{}%
\providecommand \url  [0]{\begingroup\@sanitize@url \@url }%
\providecommand \@url [1]{\endgroup\@href {#1}{\urlprefix }}%
\providecommand \urlprefix  [0]{URL }%
\providecommand \Eprint [0]{\href }%
\providecommand \doibase [0]{http://dx.doi.org/}%
\providecommand \selectlanguage [0]{\@gobble}%
\providecommand \bibinfo  [0]{\@secondoftwo}%
\providecommand \bibfield  [0]{\@secondoftwo}%
\providecommand \translation [1]{[#1]}%
\providecommand \BibitemOpen [0]{}%
\providecommand \bibitemStop [0]{}%
\providecommand \bibitemNoStop [0]{.\EOS\space}%
\providecommand \EOS [0]{\spacefactor3000\relax}%
\providecommand \BibitemShut  [1]{\csname bibitem#1\endcsname}%
\let\auto@bib@innerbib\@empty


\bibitem [{\citenamefont {{Divincenzo}}(2000)}]{Divincenzo}%
  \BibitemOpen
  \bibfield  {author} {\bibinfo {author} {\bibfnamefont {D.~P.}\ \bibnamefont
  {{Divincenzo}}},\ }\bibfield  {title} {\enquote {\bibinfo {title} {{The
  Physical Implementation of Quantum Computation}},}\ }\href@noop {} {\bibfield
   {journal} {\bibinfo  {journal} {Fortsch. Phys.}\ }\textbf {\bibinfo {volume}
  {48}},\ \bibinfo {pages} {771} (\bibinfo {year} {2000})}\BibitemShut
  {NoStop}%
\bibitem [{\citenamefont {Kimble}(2008)}]{Kimble2008}%
  \BibitemOpen
  \bibfield  {author} {\bibinfo {author} {\bibfnamefont {H.~J.}\ \bibnamefont
  {Kimble}},\ }\bibfield  {title} {\enquote {\bibinfo {title} {The quantum
  internet},}\ }\href@noop {} {\bibfield  {journal} {\bibinfo  {journal}
  {Nature}\ }\textbf {\bibinfo {volume} {453}},\ \bibinfo {pages} {1023}
  (\bibinfo {year} {2008})}\BibitemShut {NoStop}%
\bibitem [{\citenamefont {Cirac}\ \emph {et~al.}(1997)\citenamefont {Cirac},
  \citenamefont {Zoller}, \citenamefont {Kimble},\ and\ \citenamefont
  {Mabuchi}}]{Cirac1997}%
  \BibitemOpen
  \bibfield  {author} {\bibinfo {author} {\bibfnamefont {J.~I.}\ \bibnamefont
  {Cirac}}, \bibinfo {author} {\bibfnamefont {P.}~\bibnamefont {Zoller}},
  \bibinfo {author} {\bibfnamefont {H.~J.}\ \bibnamefont {Kimble}}, \ and\
  \bibinfo {author} {\bibfnamefont {H.}~\bibnamefont {Mabuchi}},\ }\bibfield
  {title} {\enquote {\bibinfo {title} {Quantum state transfer and entanglement
  distribution among distant nodes in a quantum network},}\ }\href@noop {}
  {\bibfield  {journal} {\bibinfo  {journal} {Phys. Rev. Lett.}\ }\textbf
  {\bibinfo {volume} {78}},\ \bibinfo {pages} {3221} (\bibinfo {year}
  {1997})}\BibitemShut {NoStop}%
\bibitem [{\citenamefont {Serafini}\ \emph {et~al.}(2006)\citenamefont
  {Serafini}, \citenamefont {Mancini},\ and\ \citenamefont
  {Bose}}]{Serafini2006}%
  \BibitemOpen
  \bibfield  {author} {\bibinfo {author} {\bibfnamefont {A.}~\bibnamefont
  {Serafini}}, \bibinfo {author} {\bibfnamefont {S.}~\bibnamefont {Mancini}}, \
  and\ \bibinfo {author} {\bibfnamefont {S.}~\bibnamefont {Bose}},\ }\bibfield
  {title} {\enquote {\bibinfo {title} {Distributed quantum computation via
  optical fibers},}\ }\href@noop {} {\bibfield  {journal} {\bibinfo  {journal}
  {Phys. Rev. Lett.}\ }\textbf {\bibinfo {volume} {96}},\ \bibinfo {pages}
  {010503} (\bibinfo {year} {2006})}\BibitemShut {NoStop}%
\bibitem [{\citenamefont {Yin}\ and\ \citenamefont {Li}(2007)}]{Yin2007}%
  \BibitemOpen
  \bibfield  {author} {\bibinfo {author} {\bibfnamefont {Z.-q.}\ \bibnamefont
  {Yin}}\ and\ \bibinfo {author} {\bibfnamefont {F.-l.}\ \bibnamefont {Li}},\
  }\bibfield  {title} {\enquote {\bibinfo {title} {Multiatom and resonant
  interaction scheme for quantum state transfer and logical gates between two
  remote cavities via an optical fiber},}\ }\href@noop {} {\bibfield  {journal}
  {\bibinfo  {journal} {Phys. Rev. A}\ }\textbf {\bibinfo {volume} {75}},\
  \bibinfo {pages} {012324} (\bibinfo {year} {2007})}\BibitemShut {NoStop}%
\bibitem [{\citenamefont {Axline}\ \emph {et~al.}(2018)\citenamefont {Axline},
  \citenamefont {Burkhart}, \citenamefont {Pfaff}, \citenamefont {Zhang},
  \citenamefont {Chou}, \citenamefont {Campagne-Ibarcq}, \citenamefont
  {Reinhold}, \citenamefont {Frunzio}, \citenamefont {Girvin}, \citenamefont
  {Jiang}, \citenamefont {Devoret},\ and\ \citenamefont
  {Schoelkopf}}]{Axline2018}%
  \BibitemOpen
  \bibfield  {author} {\bibinfo {author} {\bibfnamefont {C.}~\bibnamefont
  {Axline}}, \bibinfo {author} {\bibfnamefont {L.}~\bibnamefont {Burkhart}},
  \bibinfo {author} {\bibfnamefont {W.}~\bibnamefont {Pfaff}}, \bibinfo
  {author} {\bibfnamefont {M.}~\bibnamefont {Zhang}}, \bibinfo {author}
  {\bibfnamefont {K.}~\bibnamefont {Chou}}, \bibinfo {author} {\bibfnamefont
  {P.}~\bibnamefont {Campagne-Ibarcq}}, \bibinfo {author} {\bibfnamefont
  {P.}~\bibnamefont {Reinhold}}, \bibinfo {author} {\bibfnamefont
  {L.}~\bibnamefont {Frunzio}}, \bibinfo {author} {\bibfnamefont {S.~M.}\
  \bibnamefont {Girvin}}, \bibinfo {author} {\bibfnamefont {L.}~\bibnamefont
  {Jiang}}, \bibinfo {author} {\bibfnamefont {M.~H.}\ \bibnamefont {Devoret}},
  \ and\ \bibinfo {author} {\bibfnamefont {R.~J.}\ \bibnamefont {Schoelkopf}},\
  }\bibfield  {title} {\enquote {\bibinfo {title} {On-demand quantum state
  transfer and entanglement between remote microwave cavity memories},}\
  }\href@noop {} {\bibfield  {journal} {\bibinfo  {journal} {Nat. Phys.}\ }
  \textbf {\bibinfo {volume} {14}},\
  \bibinfo {pages} {705} (\bibinfo {year} {2018})}\BibitemShut {NoStop}%
\bibitem [{\citenamefont {Kurpiers}\ \emph {et~al.}(2017)\citenamefont
  {Kurpiers}, \citenamefont {Magnard}, \citenamefont {Walter}, \citenamefont
  {Royer}, \citenamefont {Pechal}, \citenamefont {Heinsoo}, \citenamefont
  {Salath{\'e}}, \citenamefont {Akin}, \citenamefont {Storz}, \citenamefont
  {Besse} \emph {et~al.}}]{Kurpiers2017}%
  \BibitemOpen
  \bibfield  {author} {\bibinfo {author} {\bibfnamefont {P.}~\bibnamefont
  {Kurpiers}}, \bibinfo {author} {\bibfnamefont {P.}~\bibnamefont {Magnard}},
  \bibinfo {author} {\bibfnamefont {T.}~\bibnamefont {Walter}}, \bibinfo
  {author} {\bibfnamefont {B.}~\bibnamefont {Royer}}, \bibinfo {author}
  {\bibfnamefont {M.}~\bibnamefont {Pechal}}, \bibinfo {author} {\bibfnamefont
  {J.}~\bibnamefont {Heinsoo}}, \bibinfo {author} {\bibfnamefont
  {Y.}~\bibnamefont {Salath{\'e}}}, \bibinfo {author} {\bibfnamefont
  {A.}~\bibnamefont {Akin}}, \bibinfo {author} {\bibfnamefont {S.}~\bibnamefont
  {Storz}}, \bibinfo {author} {\bibfnamefont {J.-C.}\ \bibnamefont {Besse}},
  \bibinfo {author} {\bibfnamefont {S.}\ \bibnamefont {Gasparinetti}},  \bibinfo {author} {\bibfnamefont {A.}\ \bibnamefont {Blais}}, \ and\ \bibinfo {author} {\bibfnamefont {A.}\ \bibnamefont {Wallraff}},\ }\bibfield  {title} {\enquote {\bibinfo {title}
  {Deterministic quantum state transfer and generation of remote entanglement
  using microwave photons},}\ }\href@noop {} {\bibfield  {journal} {\bibinfo
  {journal} {Nature}\ }   \textbf {\bibinfo {volume} {558}},\
  \bibinfo {pages} {264} (\bibinfo {year} {2018})}\BibitemShut
  {NoStop}%
\bibitem [{\citenamefont {Nielsen}\ and\ \citenamefont
  {Chuang}(2000)}]{Nielsen}%
  \BibitemOpen
  \bibfield  {author} {\bibinfo {author} {\bibfnamefont {M.~A.}\ \bibnamefont
  {Nielsen}}\ and\ \bibinfo {author} {\bibfnamefont {I.~L.}\ \bibnamefont
  {Chuang}},\ }\href@noop {} {\emph {\bibinfo {title} {Quantum Computation and
  Quantum Information}}}\ (\bibinfo  {publisher} {Cambridge Univ. Press, Cambridge},\
  \bibinfo {year} {2000})\BibitemShut {NoStop}%
\bibitem [{\citenamefont {Kielpinski}\ \emph {et~al.}(2002)\citenamefont
  {Kielpinski}, \citenamefont {Monroe},\ and\ \citenamefont
  {Wineland}}]{Kielpinski2002}%
  \BibitemOpen
  \bibfield  {author} {\bibinfo {author} {\bibfnamefont {D.}~\bibnamefont
  {Kielpinski}}, \bibinfo {author} {\bibfnamefont {C.}~\bibnamefont {Monroe}},
  \ and\ \bibinfo {author} {\bibfnamefont {D.~J.}\ \bibnamefont {Wineland}},\
  }\bibfield  {title} {\enquote {\bibinfo {title} {Architecture for a
  large-scale ion-trap quantum computer},}\ }\href@noop {} {\bibfield
  {journal} {\bibinfo  {journal} {Nature}\ }\textbf {\bibinfo {volume} {417}},\
  \bibinfo {pages} {709} (\bibinfo {year} {2002})}\BibitemShut {NoStop}%
\bibitem [{\citenamefont {Bose}(2003)}]{Bose2003}%
  \BibitemOpen
  \bibfield  {author} {\bibinfo {author} {\bibfnamefont {S.}~\bibnamefont
  {Bose}},\ }\bibfield  {title} {\enquote {\bibinfo {title} {Quantum
  communication through an unmodulated spin chain},}\ }\href@noop {} {\bibfield
   {journal} {\bibinfo  {journal} {Phys. Rev. Lett.}\ }\textbf {\bibinfo
  {volume} {91}},\ \bibinfo {pages} {207901} (\bibinfo {year}
  {2003})}\BibitemShut {NoStop}%
\bibitem [{\citenamefont {Christandl}\ \emph {et~al.}(2004)\citenamefont
  {Christandl}, \citenamefont {Datta}, \citenamefont {Ekert},\ and\
  \citenamefont {Landahl}}]{qst2004}%
  \BibitemOpen
  \bibfield  {author} {\bibinfo {author} {\bibfnamefont {M.}~\bibnamefont
  {Christandl}}, \bibinfo {author} {\bibfnamefont {N.}~\bibnamefont {Datta}},
  \bibinfo {author} {\bibfnamefont {A.}~\bibnamefont {Ekert}}, \ and\ \bibinfo
  {author} {\bibfnamefont {A.~J.}\ \bibnamefont {Landahl}},\ }\bibfield
  {title} {\enquote {\bibinfo {title} {Perfect state transfer in quantum spin
  networks},}\ }\href@noop {} {\bibfield  {journal} {\bibinfo  {journal} {Phys.
  Rev. Lett.}\ }\textbf {\bibinfo {volume} {92}},\ \bibinfo {pages} {187902}
  (\bibinfo {year} {2004})}\BibitemShut {NoStop}%
\bibitem [{\citenamefont {Romito}\ \emph {et~al.}(2005)\citenamefont {Romito},
  \citenamefont {Fazio},\ and\ \citenamefont {Bruder}}]{Romito2005}%
  \BibitemOpen
  \bibfield  {author} {\bibinfo {author} {\bibfnamefont {A.}~\bibnamefont
  {Romito}}, \bibinfo {author} {\bibfnamefont {R.}~\bibnamefont {Fazio}}, \
  and\ \bibinfo {author} {\bibfnamefont {C.}~\bibnamefont {Bruder}},\
  }\bibfield  {title} {\enquote {\bibinfo {title} {Solid-state quantum
  communication with josephson arrays},}\ }\href@noop {} {\bibfield  {journal}
  {\bibinfo  {journal} {Phys. Rev. B}\ }\textbf {\bibinfo {volume} {71}},\
  \bibinfo {pages} {100501} (\bibinfo {year} {2005})}\BibitemShut {NoStop}%
\bibitem [{\citenamefont {Yung}\ and\ \citenamefont {Bose}(2005)}]{Yung2005}%
  \BibitemOpen
  \bibfield  {author} {\bibinfo {author} {\bibfnamefont {M.-H.}\ \bibnamefont
  {Yung}}\ and\ \bibinfo {author} {\bibfnamefont {S.}~\bibnamefont {Bose}},\
  }\bibfield  {title} {\enquote {\bibinfo {title} {Perfect state transfer,
  effective gates, and entanglement generation in engineered bosonic and
  fermionic networks},}\ }\href@noop {} {\bibfield  {journal} {\bibinfo
  {journal} {Phys. Rev. A}\ }\textbf {\bibinfo {volume} {71}},\ \bibinfo
  {pages} {032310} (\bibinfo {year} {2005})}\BibitemShut {NoStop}%
\bibitem [{\citenamefont {Shi}\ \emph {et~al.}(2005)\citenamefont {Shi},
  \citenamefont {Li}, \citenamefont {Song},\ and\ \citenamefont
  {Sun}}]{Shi2005}%
  \BibitemOpen
  \bibfield  {author} {\bibinfo {author} {\bibfnamefont {T.}~\bibnamefont
  {Shi}}, \bibinfo {author} {\bibfnamefont {Y.}~\bibnamefont {Li}}, \bibinfo
  {author} {\bibfnamefont {Z.}~\bibnamefont {Song}}, \ and\ \bibinfo {author}
  {\bibfnamefont {C.-P.}\ \bibnamefont {Sun}},\ }\bibfield  {title} {\enquote
  {\bibinfo {title} {Quantum-state transfer via the ferromagnetic chain in a
  spatially modulated field},}\ }\href@noop {} {\bibfield  {journal} {\bibinfo
  {journal} {Phys. Rev. A}\ }\textbf {\bibinfo {volume} {71}},\ \bibinfo
  {pages} {032309} (\bibinfo {year} {2005})}\BibitemShut {NoStop}%
\bibitem [{\citenamefont {Bose}(2007)}]{Bose2007}%
  \BibitemOpen
  \bibfield  {author} {\bibinfo {author} {\bibfnamefont {S.}~\bibnamefont
  {Bose}},\ }\bibfield  {title} {\enquote {\bibinfo {title} {Quantum
  communication through spin chain dynamics: an introductory overview},}\
  }\href@noop {} {\bibfield  {journal} {\bibinfo  {journal} {Contemp. Phys.}\
  }\textbf {\bibinfo {volume} {48}},\ \bibinfo {pages} {13} (\bibinfo {year}
  {2007})}\BibitemShut {NoStop}%
\bibitem [{\citenamefont {Di~Franco}\ \emph {et~al.}(2008)\citenamefont
  {Di~Franco}, \citenamefont {Paternostro},\ and\ \citenamefont
  {Kim}}]{Franco2008}%
  \BibitemOpen
  \bibfield  {author} {\bibinfo {author} {\bibfnamefont {C.}~\bibnamefont
  {Di~Franco}}, \bibinfo {author} {\bibfnamefont {M.}~\bibnamefont
  {Paternostro}}, \ and\ \bibinfo {author} {\bibfnamefont {M. S.}~\bibnamefont
  {Kim}},\ }\bibfield  {title} {\enquote {\bibinfo {title} {Perfect state
  transfer on a spin chain without state initialization},}\ }\href@noop {}
  {\bibfield  {journal} {\bibinfo  {journal} {Phys. Rev. Lett.}\ }\textbf
  {\bibinfo {volume} {101}},\ \bibinfo {pages} {230502} (\bibinfo {year}
  {2008})}\BibitemShut {NoStop}%
\bibitem [{\citenamefont {Yao}\ \emph {et~al.}(2011)\citenamefont {Yao},
  \citenamefont {Jiang}, \citenamefont {Gorshkov}, \citenamefont {Gong},
  \citenamefont {Zhai}, \citenamefont {Duan},\ and\ \citenamefont
  {Lukin}}]{Yao2011}%
  \BibitemOpen
  \bibfield  {author} {\bibinfo {author} {\bibfnamefont {N.~Y.}\ \bibnamefont
  {Yao}}, \bibinfo {author} {\bibfnamefont {L.}~\bibnamefont {Jiang}}, \bibinfo
  {author} {\bibfnamefont {A.~V.}\ \bibnamefont {Gorshkov}}, \bibinfo {author}
  {\bibfnamefont {Z.-X.}\ \bibnamefont {Gong}}, \bibinfo {author}
  {\bibfnamefont {A.}~\bibnamefont {Zhai}}, \bibinfo {author} {\bibfnamefont
  {L.-M.}\ \bibnamefont {Duan}}, \ and\ \bibinfo {author} {\bibfnamefont
  {M.~D.}\ \bibnamefont {Lukin}},\ }\bibfield  {title} {\enquote {\bibinfo
  {title} {Robust quantum state transfer in random unpolarized spin chains},}\
  }\href@noop {} {\bibfield  {journal} {\bibinfo  {journal} {Phys. Rev. Lett.}\
  }\textbf {\bibinfo {volume} {106}},\ \bibinfo {pages} {040505} (\bibinfo
  {year} {2011})}\BibitemShut {NoStop}%
\bibitem [{\citenamefont {Zhang}\ \emph {et~al.}(2005)\citenamefont {Zhang},
  \citenamefont {Long}, \citenamefont {Zhang}, \citenamefont {Deng},
  \citenamefont {Liu},\ and\ \citenamefont {Lu}}]{Zhang2005}%
  \BibitemOpen
  \bibfield  {author} {\bibinfo {author} {\bibfnamefont {J.}~\bibnamefont
  {Zhang}}, \bibinfo {author} {\bibfnamefont {G.~L.}\ \bibnamefont {Long}},
  \bibinfo {author} {\bibfnamefont {W.}~\bibnamefont {Zhang}}, \bibinfo
  {author} {\bibfnamefont {Z.}~\bibnamefont {Deng}}, \bibinfo {author}
  {\bibfnamefont {W.}~\bibnamefont {Liu}}, \ and\ \bibinfo {author}
  {\bibfnamefont {Z.}~\bibnamefont {Lu}},\ }\bibfield  {title} {\enquote
  {\bibinfo {title} {Simulation of heisenberg $xy$ interactions and realization
  of a perfect state transfer in spin chains using liquid nuclear magnetic
  resonance},}\ }\href@noop {} {\bibfield  {journal} {\bibinfo  {journal}
  {Phys. Rev. A}\ }\textbf {\bibinfo {volume} {72}},\ \bibinfo {pages} {012331}
  (\bibinfo {year} {2005})}\BibitemShut {NoStop}%
\bibitem [{\citenamefont {Perez-Leija}\ \emph {et~al.}(2013)\citenamefont
  {Perez-Leija}, \citenamefont {Keil}, \citenamefont {Kay}, \citenamefont
  {Moya-Cessa}, \citenamefont {Nolte}, \citenamefont {Kwek}, \citenamefont
  {Rodr\'{\i}guez-Lara}, \citenamefont {Szameit},\ and\ \citenamefont
  {Christodoulides}}]{Perez2013}%
  \BibitemOpen
  \bibfield  {author} {\bibinfo {author} {\bibfnamefont {A.}~\bibnamefont
  {Perez-Leija}}, \bibinfo {author} {\bibfnamefont {R.}~\bibnamefont {Keil}},
  \bibinfo {author} {\bibfnamefont {A.}~\bibnamefont {Kay}}, \bibinfo {author}
  {\bibfnamefont {H.}~\bibnamefont {Moya-Cessa}}, \bibinfo {author}
  {\bibfnamefont {S.}~\bibnamefont {Nolte}}, \bibinfo {author} {\bibfnamefont
  {L.-C.}\ \bibnamefont {Kwek}}, \bibinfo {author} {\bibfnamefont {B.~M.}\
  \bibnamefont {Rodr\'{\i}guez-Lara}}, \bibinfo {author} {\bibfnamefont
  {A.}~\bibnamefont {Szameit}}, \ and\ \bibinfo {author} {\bibfnamefont
  {D.~N.}\ \bibnamefont {Christodoulides}},\ }\bibfield  {title} {\enquote
  {\bibinfo {title} {Coherent quantum transport in photonic lattices},}\
  }\href@noop {} {\bibfield  {journal} {\bibinfo  {journal} {Phys. Rev. A}\
  }\textbf {\bibinfo {volume} {87}},\ \bibinfo {pages} {012309} (\bibinfo
  {year} {2013})}\BibitemShut {NoStop}%
\bibitem [{\citenamefont {Chapman}\ \emph {et~al.}(2016)\citenamefont
  {Chapman}, \citenamefont {Santandrea}, \citenamefont {Huang}, \citenamefont
  {Corrielli}, \citenamefont {Crespi}, \citenamefont {Yung}, \citenamefont
  {Osellame},\ and\ \citenamefont {Peruzzo}}]{Chapman2016}%
  \BibitemOpen
  \bibfield  {author} {\bibinfo {author} {\bibfnamefont {R.~J.}\ \bibnamefont
  {Chapman}}, \bibinfo {author} {\bibfnamefont {M.}~\bibnamefont {Santandrea}},
  \bibinfo {author} {\bibfnamefont {Z.}~\bibnamefont {Huang}}, \bibinfo
  {author} {\bibfnamefont {G.}~\bibnamefont {Corrielli}}, \bibinfo {author}
  {\bibfnamefont {A.}~\bibnamefont {Crespi}}, \bibinfo {author} {\bibfnamefont
  {M.-H.}\ \bibnamefont {Yung}}, \bibinfo {author} {\bibfnamefont
  {R.}~\bibnamefont {Osellame}}, \ and\ \bibinfo {author} {\bibfnamefont
  {A.}~\bibnamefont {Peruzzo}},\ }\bibfield  {title} {\enquote {\bibinfo
  {title} {Experimental perfect state transfer of an entangled photonic
  qubit},}\ }\href@noop {} {\bibfield  {journal} {\bibinfo  {journal} {Nat.
  Commun.}\ }\textbf {\bibinfo {volume} {7}},\ \bibinfo {pages} {11339} (\bibinfo
  {year} {2016})}\BibitemShut {NoStop}%
\bibitem [{\citenamefont {Chow}\ \emph {et~al.}(2011)\citenamefont {Chow},
  \citenamefont {C\'orcoles}, \citenamefont {Gambetta}, \citenamefont
  {Rigetti}, \citenamefont {Johnson}, \citenamefont {Smolin}, \citenamefont
  {Rozen}, \citenamefont {Keefe}, \citenamefont {Rothwell}, \citenamefont
  {Ketchen},\ and\ \citenamefont {Steffen}}]{Chow2011}%
  \BibitemOpen
  \bibfield  {author} {\bibinfo {author} {\bibfnamefont {J.~M.}\ \bibnamefont
  {Chow}}, \bibinfo {author} {\bibfnamefont {A.~D.}\ \bibnamefont
  {C\'orcoles}}, \bibinfo {author} {\bibfnamefont {J.~M.}\ \bibnamefont
  {Gambetta}}, \bibinfo {author} {\bibfnamefont {C.}~\bibnamefont {Rigetti}},
  \bibinfo {author} {\bibfnamefont {B.~R.}\ \bibnamefont {Johnson}}, \bibinfo
  {author} {\bibfnamefont {J.~A.}\ \bibnamefont {Smolin}}, \bibinfo {author}
  {\bibfnamefont {J.~R.}\ \bibnamefont {Rozen}}, \bibinfo {author}
  {\bibfnamefont {G.~A.}\ \bibnamefont {Keefe}}, \bibinfo {author}
  {\bibfnamefont {M.~B.}\ \bibnamefont {Rothwell}}, \bibinfo {author}
  {\bibfnamefont {M.~B.}\ \bibnamefont {Ketchen}}, \ and\ \bibinfo {author}
  {\bibfnamefont {M.}~\bibnamefont {Steffen}},\ }\bibfield  {title} {\enquote
  {\bibinfo {title} {Simple all-microwave entangling gate for fixed-frequency
  superconducting qubits},}\ }\href@noop {} {\bibfield  {journal} {\bibinfo
  {journal} {Phys. Rev. Lett.}\ }\textbf {\bibinfo {volume} {107}},\ \bibinfo
  {pages} {080502} (\bibinfo {year} {2011})}\BibitemShut {NoStop}%
\bibitem [{\citenamefont {Poletto}\ \emph {et~al.}(2012)\citenamefont
  {Poletto}, \citenamefont {Gambetta}, \citenamefont {Merkel}, \citenamefont
  {Smolin}, \citenamefont {Chow}, \citenamefont {C\'orcoles}, \citenamefont
  {Keefe}, \citenamefont {Rothwell}, \citenamefont {Rozen}, \citenamefont
  {Abraham}, \citenamefont {Rigetti},\ and\ \citenamefont
  {Steffen}}]{Poletto2012}%
  \BibitemOpen
  \bibfield  {author} {\bibinfo {author} {\bibfnamefont {S.}~\bibnamefont
  {Poletto}}, \bibinfo {author} {\bibfnamefont {J.~M.}\ \bibnamefont
  {Gambetta}}, \bibinfo {author} {\bibfnamefont {S.~T.}\ \bibnamefont
  {Merkel}}, \bibinfo {author} {\bibfnamefont {J.~A.}\ \bibnamefont {Smolin}},
  \bibinfo {author} {\bibfnamefont {J.~M.}\ \bibnamefont {Chow}}, \bibinfo
  {author} {\bibfnamefont {A.~D.}\ \bibnamefont {C\'orcoles}}, \bibinfo
  {author} {\bibfnamefont {G.~A.}\ \bibnamefont {Keefe}}, \bibinfo {author}
  {\bibfnamefont {M.~B.}\ \bibnamefont {Rothwell}}, \bibinfo {author}
  {\bibfnamefont {J.~R.}\ \bibnamefont {Rozen}}, \bibinfo {author}
  {\bibfnamefont {D.~W.}\ \bibnamefont {Abraham}}, \bibinfo {author}
  {\bibfnamefont {C.}~\bibnamefont {Rigetti}}, \ and\ \bibinfo {author}
  {\bibfnamefont {M.}~\bibnamefont {Steffen}},\ }\bibfield  {title} {\enquote
  {\bibinfo {title} {Entanglement of two superconducting qubits in a waveguide
  cavity via monochromatic two-photon excitation},}\ }\href@noop {} {\bibfield
  {journal} {\bibinfo  {journal} {Phys. Rev. Lett.}\ }\textbf {\bibinfo
  {volume} {109}},\ \bibinfo {pages} {240505} (\bibinfo {year}
  {2012})}\BibitemShut {NoStop}%
\bibitem [{\citenamefont {DiCarlo}\ \emph {et~al.}(2009)\citenamefont
  {DiCarlo}, \citenamefont {Chow}, \citenamefont {Gambetta}, \citenamefont
  {Bishop}, \citenamefont {Johnson}, \citenamefont {Schuster}, \citenamefont
  {Majer}, \citenamefont {Blais}, \citenamefont {Frunzio}, \citenamefont
  {Girvin},\ and\ \citenamefont {Schoelkopf}}]{DiCarlo2009}%
  \BibitemOpen
  \bibfield  {author} {\bibinfo {author} {\bibfnamefont {L.}~\bibnamefont
  {DiCarlo}}, \bibinfo {author} {\bibfnamefont {J.~M.}\ \bibnamefont {Chow}},
  \bibinfo {author} {\bibfnamefont {J.~M.}\ \bibnamefont {Gambetta}}, \bibinfo
  {author} {\bibfnamefont {L.~S.}\ \bibnamefont {Bishop}}, \bibinfo {author}
  {\bibfnamefont {B.~R.}\ \bibnamefont {Johnson}}, \bibinfo {author}
  {\bibfnamefont {D.~I.}\ \bibnamefont {Schuster}}, \bibinfo {author}
  {\bibfnamefont {J.}~\bibnamefont {Majer}}, \bibinfo {author} {\bibfnamefont
  {A.}~\bibnamefont {Blais}}, \bibinfo {author} {\bibfnamefont
  {L.}~\bibnamefont {Frunzio}}, \bibinfo {author} {\bibfnamefont {S.~M.}\
  \bibnamefont {Girvin}}, \ and\ \bibinfo {author} {\bibfnamefont {R.~J.}\
  \bibnamefont {Schoelkopf}},\ }\bibfield  {title} {\enquote {\bibinfo {title}
  {Demonstration of two-qubit algorithms with a superconducting quantum
  processor},}\ }\href@noop {} {\bibfield  {journal} {\bibinfo  {journal}
  {Nature}\ }\textbf {\bibinfo {volume} {460}},\ \bibinfo {pages} {240}
  (\bibinfo {year} {2009})}\BibitemShut {NoStop}%
\bibitem [{\citenamefont {Kelly}\ \emph {et~al.}(2015)\citenamefont {Kelly},
  \citenamefont {Barends}, \citenamefont {Fowler}, \citenamefont {Megrant},
  \citenamefont {Jeffrey}, \citenamefont {White}, \citenamefont {Sank},
  \citenamefont {Mutus}, \citenamefont {Campbell}, \citenamefont {Chen},
  \citenamefont {Chen}, \citenamefont {Chiaro}, \citenamefont {Dunsworth},
  \citenamefont {Hoi}, \citenamefont {Neill}, \citenamefont {O'Malley},
  \citenamefont {Quintana}, \citenamefont {Roushan}, \citenamefont
  {Vainsencher}, \citenamefont {Wenner}, \citenamefont {Cleland},\ and\
  \citenamefont {Martinis}}]{Kelly2015}%
  \BibitemOpen
\bibfield  {author} {\bibinfo {author} {\bibfnamefont {J.}~\bibnamefont   {Kelly}}, \bibinfo {author} {\bibfnamefont {R.}~\bibnamefont {Barends}},   \bibinfo {author} {\bibfnamefont {A.~G.}\ \bibnamefont {Fowler}}, \bibinfo   {author} {\bibfnamefont {A.}~\bibnamefont {Megrant}}, \bibinfo {author}   {\bibfnamefont {E.}~\bibnamefont {Jeffrey}}, \bibinfo {author} {\bibfnamefont   {T.~C.}\ \bibnamefont {White}}, \bibinfo {author} {\bibfnamefont   {D.}~\bibnamefont {Sank}}, \bibinfo {author} {\bibfnamefont {J.~Y.}\   \bibnamefont {Mutus}}, \bibinfo {author} {\bibfnamefont {B.}~\bibnamefont   {Campbell}}, \bibinfo {author} {\bibfnamefont {Y.}~\bibnamefont {Chen}},
%\bibinfo {author} {\bibfnamefont {Z.}~\bibnamefont {Chen}}, \bibinfo {author}   {\bibfnamefont {B.}~\bibnamefont {Chiaro}}, \bibinfo {author} {\bibfnamefont   {A.}~\bibnamefont {Dunsworth}}, \bibinfo {author} {\bibfnamefont {I.-C.}\   \bibnamefont {Hoi}}, \bibinfo {author} {\bibfnamefont {C.}~\bibnamefont   {Neill}}, \bibinfo {author} {\bibfnamefont {P.~J.~J.}\ \bibnamefont   {O'Malley}}, \bibinfo {author} {\bibfnamefont {C.}~\bibnamefont {Quintana}},   \bibinfo {author} {\bibfnamefont {P.}~\bibnamefont {Roushan}}, \bibinfo   {author} {\bibfnamefont {A.}~\bibnamefont {Vainsencher}}, \bibinfo {author}   {\bibfnamefont {J.}~\bibnamefont {Wenner}}, \bibinfo {author} {\bibfnamefont   {A.~N.}\ \bibnamefont {Cleland}}, \ and\ \bibinfo {author} {\bibfnamefont   {J.~M.}\ \bibnamefont {Martinis}},
  \emph {et~al.}, \ }\bibfield  {title} {\enquote {\bibinfo
  {title} {{State preservation by repetitive error detection in a
  superconducting quantum circuit}},}\ }\href@noop {} {\bibfield  {journal}
  {\bibinfo  {journal} {Nature}\ }\textbf {\bibinfo {volume} {519}},\ \bibinfo
  {pages} {66} (\bibinfo {year} {2015})}\BibitemShut {NoStop}%
\bibitem [{\citenamefont {Liu}\ \emph {et~al.}(2006)\citenamefont {Liu},
  \citenamefont {Wei}, \citenamefont {Tsai},\ and\ \citenamefont
  {Nori}}]{Liu2006}%
  \BibitemOpen
  \bibfield  {author} {\bibinfo {author} {\bibfnamefont {Y.-X.}\ \bibnamefont
  {Liu}}, \bibinfo {author} {\bibfnamefont {L.~F.}\ \bibnamefont {Wei}},
  \bibinfo {author} {\bibfnamefont {J.~S.}\ \bibnamefont {Tsai}}, \ and\
  \bibinfo {author} {\bibfnamefont {F.}~\bibnamefont {Nori}},\ }\bibfield
  {title} {\enquote {\bibinfo {title} {Controllable coupling between flux
  qubits},}\ }\href@noop {} {\bibfield  {journal} {\bibinfo  {journal} {Phys.
  Rev. Lett.}\ }\textbf {\bibinfo {volume} {96}},\ \bibinfo {pages} {067003}
  (\bibinfo {year} {2006})}\BibitemShut {NoStop}%
\bibitem [{\citenamefont {Niskanen}\ \emph {et~al.}(2007)\citenamefont
  {Niskanen}, \citenamefont {Harrabi}, \citenamefont {Yoshihara}, \citenamefont
  {Nakamura}, \citenamefont {Lloyd},\ and\ \citenamefont
  {Tsai}}]{Niskanen2007}%
  \BibitemOpen
  \bibfield  {author} {\bibinfo {author} {\bibfnamefont {A.}~\bibnamefont
  {Niskanen}}, \bibinfo {author} {\bibfnamefont {K.}~\bibnamefont {Harrabi}},
  \bibinfo {author} {\bibfnamefont {F.}~\bibnamefont {Yoshihara}}, \bibinfo
  {author} {\bibfnamefont {Y.}~\bibnamefont {Nakamura}}, \bibinfo {author}
  {\bibfnamefont {S.}~\bibnamefont {Lloyd}}, \ and\ \bibinfo {author}
  {\bibfnamefont {J.}~\bibnamefont {Tsai}},\ }\bibfield  {title} {\enquote
  {\bibinfo {title} {Quantum coherent tunable coupling of superconducting
  qubits},}\ }\href@noop {} {\bibfield  {journal} {\bibinfo  {journal}
  {Science}\ }\textbf {\bibinfo {volume} {316}},\ \bibinfo {pages} {723}
  (\bibinfo {year} {2007})}\BibitemShut {NoStop}%
\bibitem [{\citenamefont {McKay}\ \emph {et~al.}(2016)\citenamefont {McKay},
  \citenamefont {Filipp}, \citenamefont {Mezzacapo}, \citenamefont {Magesan},
  \citenamefont {Chow},\ and\ \citenamefont {Gambetta}}]{McKay2016}%
  \BibitemOpen
  \bibfield  {author} {\bibinfo {author} {\bibfnamefont {D.~C.}\ \bibnamefont
  {McKay}}, \bibinfo {author} {\bibfnamefont {S.}~\bibnamefont {Filipp}},
  \bibinfo {author} {\bibfnamefont {A.}~\bibnamefont {Mezzacapo}}, \bibinfo
  {author} {\bibfnamefont {E.}~\bibnamefont {Magesan}}, \bibinfo {author}
  {\bibfnamefont {J.~M.}\ \bibnamefont {Chow}}, \ and\ \bibinfo {author}
  {\bibfnamefont {J.~M.}\ \bibnamefont {Gambetta}},\ }\bibfield  {title}
  {\enquote {\bibinfo {title} {Universal gate for fixed-frequency qubits via a
  tunable bus},}\ }\href@noop {} {\bibfield  {journal} {\bibinfo  {journal}
  {Phys. Rev. Appl.}\ }\textbf {\bibinfo {volume} {6}},\ \bibinfo {pages}
  {064007} (\bibinfo {year} {2016})}\BibitemShut {NoStop}%
\bibitem [{\citenamefont {Naik}\ \emph {et~al.}(2017)\citenamefont {Naik},
  \citenamefont {Leung}, \citenamefont {Chakram}, \citenamefont {Groszkowski},
  \citenamefont {Lu}, \citenamefont {Earnest}, \citenamefont {McKay},
  \citenamefont {Koch},\ and\ \citenamefont {Schuster}}]{Naik2017}%
  \BibitemOpen
  \bibfield  {author} {\bibinfo {author} {\bibfnamefont {R.}~\bibnamefont
  {Naik}}, \bibinfo {author} {\bibfnamefont {N.}~\bibnamefont {Leung}},
  \bibinfo {author} {\bibfnamefont {S.}~\bibnamefont {Chakram}}, \bibinfo
  {author} {\bibfnamefont {P.}~\bibnamefont {Groszkowski}}, \bibinfo {author}
  {\bibfnamefont {Y.}~\bibnamefont {Lu}}, \bibinfo {author} {\bibfnamefont
  {N.}~\bibnamefont {Earnest}}, \bibinfo {author} {\bibfnamefont
  {D.}~\bibnamefont {McKay}}, \bibinfo {author} {\bibfnamefont
  {J.}~\bibnamefont {Koch}}, \ and\ \bibinfo {author} {\bibfnamefont
  {D.}~\bibnamefont {Schuster}},\ }\bibfield  {title} {\enquote {\bibinfo
  {title} {Random access quantum information processors using multimode circuit
  quantum electrodynamics},}\ }\href@noop {} {\bibfield  {journal} {\bibinfo
  {journal} {Nat. Commun.}\ }\textbf {\bibinfo {volume} {8}},\ \bibinfo {pages}
  {1904} (\bibinfo {year} {2017})}\BibitemShut {NoStop}%
\bibitem [{\citenamefont {Lu}\ \emph {et~al.}(2017)\citenamefont {Lu},
  \citenamefont {Chakram}, \citenamefont {Leung}, \citenamefont {Earnest},
  \citenamefont {Naik}, \citenamefont {Huang}, \citenamefont {Groszkowski},
  \citenamefont {Kapit}, \citenamefont {Koch},\ and\ \citenamefont
  {Schuster}}]{Lu2017}%
  \BibitemOpen
  \bibfield  {author} {\bibinfo {author} {\bibfnamefont {Y.}~\bibnamefont
  {Lu}}, \bibinfo {author} {\bibfnamefont {S.}~\bibnamefont {Chakram}},
  \bibinfo {author} {\bibfnamefont {N.}~\bibnamefont {Leung}}, \bibinfo
  {author} {\bibfnamefont {N.}~\bibnamefont {Earnest}}, \bibinfo {author}
  {\bibfnamefont {R.~K.}\ \bibnamefont {Naik}}, \bibinfo {author}
  {\bibfnamefont {Z.}~\bibnamefont {Huang}}, \bibinfo {author} {\bibfnamefont
  {P.}~\bibnamefont {Groszkowski}}, \bibinfo {author} {\bibfnamefont
  {E.}~\bibnamefont {Kapit}}, \bibinfo {author} {\bibfnamefont
  {J.}~\bibnamefont {Koch}}, \ and\ \bibinfo {author} {\bibfnamefont {D.~I.}\
  \bibnamefont {Schuster}},\ }\bibfield  {title} {\enquote {\bibinfo {title}
  {Universal stabilization of a parametrically coupled qubit},}\ }\href@noop {}
  {\bibfield  {journal} {\bibinfo  {journal} {Phys. Rev. Lett.}\ }\textbf
  {\bibinfo {volume} {119}},\ \bibinfo {pages} {150502} (\bibinfo {year}
  {2017})}\BibitemShut {NoStop}%
\bibitem [{\citenamefont {Chen}\ \emph {et~al.}(2017)\citenamefont {Chen},
  \citenamefont {Wu}, \citenamefont {Sun},\ and\ \citenamefont
  {Liu}}]{chen2017}%
  \BibitemOpen
  \bibfield  {author} {\bibinfo {author} {\bibfnamefont {Q.~M.}\ \bibnamefont
  {Chen}}, \bibinfo {author} {\bibfnamefont {R.~B.}\ \bibnamefont {Wu}},
  \bibinfo {author} {\bibfnamefont {L.}~\bibnamefont {Sun}}, \ and\ \bibinfo
  {author} {\bibfnamefont {Y.~X.}\ \bibnamefont {Liu}},\ }\bibfield  {title}
  {\enquote {\bibinfo {title} {Tuning coupling between superconducting
  resonators with collective qubits},}\ }\href@noop {} {\bibfield  {journal}
  {\bibinfo  {journal} {arXiv:1712.04357}\ } (\bibinfo {year}
  {2017})}\BibitemShut {NoStop}%
\bibitem [{\citenamefont {{Roushan}}\ \emph {et~al.}(2017)\citenamefont
  {{Roushan}}, \citenamefont {{Neill}}, \citenamefont {{Megrant}},
  \citenamefont {{Chen}}, \citenamefont {{Babbush}}, \citenamefont {{Barends}},
  \citenamefont {{Campbell}}, \citenamefont {{Chen}}, \citenamefont {{Chiaro}},
  \citenamefont {{Dunsworth}}, \citenamefont {{Fowler}}, \citenamefont
  {{Jeffrey}}, \citenamefont {{Kelly}}, \citenamefont {{Lucero}}, \citenamefont
  {{Mutus}}, \citenamefont {{O'Malley}}, \citenamefont {{Neeley}},
  \citenamefont {{Quintana}}, \citenamefont {{Sank}}, \citenamefont
  {{Vainsencher}}, \citenamefont {{Wenner}}, \citenamefont {{White}},
  \citenamefont {{Kapit}}, \citenamefont {{Neven}},\ and\ \citenamefont
  {{Martinis}}}]{Roushan2017}%
  \BibitemOpen
\bibfield  {author} {\bibinfo {author} {\bibfnamefont {P.}~\bibnamefont   {{Roushan}}},
\bibinfo {author} {\bibfnamefont {C.}~\bibnamefont {{Neill}}},   \bibinfo {author} {\bibfnamefont {A.}~\bibnamefont {{Megrant}}}, \bibinfo   {author} {\bibfnamefont {Y.}~\bibnamefont {{Chen}}}, \bibinfo {author}   {\bibfnamefont {R.}~\bibnamefont {{Babbush}}}, \bibinfo {author}   {\bibfnamefont {R.}~\bibnamefont {{Barends}}}, \bibinfo {author}   {\bibfnamefont {B.}~\bibnamefont {{Campbell}}}, \bibinfo {author}   {\bibfnamefont {Z.}~\bibnamefont {{Chen}}}, \bibinfo {author} {\bibfnamefont   {B.}~\bibnamefont {{Chiaro}}}, \bibinfo {author} {\bibfnamefont   {A.}~\bibnamefont {{Dunsworth}}}
%, \bibinfo {author} {\bibfnamefont   {A.}~\bibnamefont {{Fowler}}}, \bibinfo {author} {\bibfnamefont   {E.}~\bibnamefont {{Jeffrey}}}, \bibinfo {author} {\bibfnamefont   {J.}~\bibnamefont {{Kelly}}}, \bibinfo {author} {\bibfnamefont  {E.}~\bibnamefont {{Lucero}}}, \bibinfo {author} {\bibfnamefont  {J.}~\bibnamefont {{Mutus}}}, \bibinfo {author} {\bibfnamefont {P.~J.~J.}\   \bibnamefont {{O'Malley}}}, \bibinfo {author} {\bibfnamefont   {M.}~\bibnamefont {{Neeley}}}, \bibinfo {author} {\bibfnamefont   {C.}~\bibnamefont {{Quintana}}}, \bibinfo {author} {\bibfnamefont   {D.}~\bibnamefont {{Sank}}}, \bibinfo {author} {\bibfnamefont   {A.}~\bibnamefont {{Vainsencher}}}, \bibinfo {author} {\bibfnamefont   {J.}~\bibnamefont {{Wenner}}}, \bibinfo {author} {\bibfnamefont   {T.}~\bibnamefont {{White}}}, \bibinfo {author} {\bibfnamefont   {E.}~\bibnamefont {{Kapit}}}, \bibinfo {author} {\bibfnamefont   {H.}~\bibnamefont {{Neven}}}, \ and\ \bibinfo {author} {\bibfnamefont   {J.}~\bibnamefont {{Martinis}}},
\emph {et~al.},\ }\bibfield  {title} {\enquote {\bibinfo
  {title} {{Chiral ground-state currents of interacting photons in a synthetic
  magnetic field}},}\ }\href@noop {} {\bibfield  {journal} {\bibinfo  {journal}
  {Nat. Phys.}\ }\textbf {\bibinfo {volume} {13}},\ \bibinfo {pages} {146}
  (\bibinfo {year} {2017})}\BibitemShut {NoStop}%
\bibitem [{\citenamefont {Neill}\ \emph {et~al.}(2018)\citenamefont {Neill},
  \citenamefont {Roushan}, \citenamefont {Kechedzhi}, \citenamefont {Boixo},
  \citenamefont {Isakov}, \citenamefont {Smelyanskiy}, \citenamefont {Megrant},
  \citenamefont {Chiaro}, \citenamefont {Dunsworth}, \citenamefont {Arya} \emph
  {et~al.}}]{Neill2018}%
  \BibitemOpen
\bibfield  {author} {\bibinfo {author} {\bibfnamefont {C.}~\bibnamefont   {Neill}}, \bibinfo {author} {\bibfnamefont {P.}~\bibnamefont {Roushan}},   \bibinfo {author} {\bibfnamefont {K.}~\bibnamefont {Kechedzhi}}, \bibinfo  {author} {\bibfnamefont {S.}~\bibnamefont {Boixo}}, \bibinfo {author}  {\bibfnamefont {S.}~\bibnamefont {Isakov}}, \bibinfo {author} {\bibfnamefont  {V.}~\bibnamefont {Smelyanskiy}}, \bibinfo {author} {\bibfnamefont  {A.}~\bibnamefont {Megrant}}, \bibinfo {author} {\bibfnamefont  {B.}~\bibnamefont {Chiaro}}, \bibinfo {author} {\bibfnamefont  {A.}~\bibnamefont {Dunsworth}}, \bibinfo {author} {\bibfnamefont  {K.}~\bibnamefont {Arya}}
\emph {et~al.},\ }\bibfield  {title} {\enquote
  {\bibinfo {title} {A blueprint for demonstrating quantum supremacy with
  superconducting qubits},}\ }\href@noop {} {\bibfield  {journal} {\bibinfo
  {journal} {Science}\ }\textbf {\bibinfo {volume} {360}},\ \bibinfo {pages}
  {195} (\bibinfo {year} {2018})}\BibitemShut {NoStop}%
\bibitem [{\citenamefont {Zhou}\ \emph {et~al.}(2009)\citenamefont {Zhou},
  \citenamefont {Yang}, \citenamefont {Liu}, \citenamefont {Sun},\ and\
  \citenamefont {Nori}}]{Zhou2009}%
  \BibitemOpen
  \bibfield  {author} {\bibinfo {author} {\bibfnamefont {L.}~\bibnamefont
  {Zhou}}, \bibinfo {author} {\bibfnamefont {S.}~\bibnamefont {Yang}}, \bibinfo
  {author} {\bibfnamefont {Y.-x.}\ \bibnamefont {Liu}}, \bibinfo {author}
  {\bibfnamefont {C.~P.}\ \bibnamefont {Sun}}, \ and\ \bibinfo {author}
  {\bibfnamefont {F.}~\bibnamefont {Nori}},\ }\bibfield  {title} {\enquote
  {\bibinfo {title} {Quantum zeno switch for single-photon coherent
  transport},}\ }\href@noop {} {\bibfield  {journal} {\bibinfo  {journal}
  {Phys. Rev. A}\ }\textbf {\bibinfo {volume} {80}},\ \bibinfo {pages} {062109}
  (\bibinfo {year} {2009})}\BibitemShut {NoStop}%
\bibitem [{\citenamefont {Strand}\ \emph {et~al.}(2013)\citenamefont {Strand},
  \citenamefont {Ware}, \citenamefont {Beaudoin}, \citenamefont {Ohki},
  \citenamefont {Johnson}, \citenamefont {Blais},\ and\ \citenamefont
  {Plourde}}]{Strand2013}%
  \BibitemOpen
  \bibfield  {author} {\bibinfo {author} {\bibfnamefont {J. D.}~\bibnamefont
  {Strand}}, \bibinfo {author} {\bibfnamefont {M.}~\bibnamefont {Ware}},
  \bibinfo {author} {\bibfnamefont {F.}~\bibnamefont {Beaudoin}}, \bibinfo
  {author} {\bibfnamefont {T. A.}~\bibnamefont {Ohki}}, \bibinfo {author}
  {\bibfnamefont {B. R.}~\bibnamefont {Johnson}}, \bibinfo {author} {\bibfnamefont
  {A.}~\bibnamefont {Blais}}, \ and\ \bibinfo {author} {\bibfnamefont
  {B. L. T.}~\bibnamefont {Plourde}},\ }\bibfield  {title} {\enquote {\bibinfo
  {title} {First-order sideband transitions with flux-driven asymmetric
  transmon qubits},}\ }\href@noop {} {\bibfield  {journal} {\bibinfo  {journal}
  {Phys. Rev. B}\ }\textbf {\bibinfo {volume} {87}},\ \bibinfo {pages} {220505}
  (\bibinfo {year} {2013})}\BibitemShut {NoStop}%
\bibitem [{\citenamefont {Liu}\ \emph {et~al.}(2014)\citenamefont {Liu},
  \citenamefont {Wang}, \citenamefont {Sun},\ and\ \citenamefont
  {Wang}}]{Liu2014}%
  \BibitemOpen
  \bibfield  {author} {\bibinfo {author} {\bibfnamefont {Y.~X.}\ \bibnamefont
  {Liu}}, \bibinfo {author} {\bibfnamefont {C.~X.}\ \bibnamefont {Wang}},
  \bibinfo {author} {\bibfnamefont {H.~C.}\ \bibnamefont {Sun}}, \ and\
  \bibinfo {author} {\bibfnamefont {X.~B.}\ \bibnamefont {Wang}},\ }\bibfield
  {title} {\enquote {\bibinfo {title} {Coexistence of single- and multi-photon
  processes due to longitudinal couplings between superconducting flux qubits
  and external fields},}\ }\href@noop {} {\bibfield  {journal} {\bibinfo
  {journal} {New J. Phys.}\ }\textbf {\bibinfo {volume} {16}},\ \bibinfo
  {pages} {015031} (\bibinfo {year} {2014})}\BibitemShut {NoStop}%
\bibitem [{\citenamefont {Wu}\ \emph {et~al.}(2016)\citenamefont {Wu},
  \citenamefont {Yang}, \citenamefont {Zheng}, \citenamefont {Deng},
  \citenamefont {Yan}, \citenamefont {Zhao}, \citenamefont {Huang},
  \citenamefont {Munro}, \citenamefont {Nemoto}, \citenamefont {Zheng},
  \citenamefont {Sun}, \citenamefont {Liu}, \citenamefont {Zhu},\ and\
  \citenamefont {Lu}}]{Wu2016}%
  \BibitemOpen
  \bibfield  {author} {\bibinfo {author} {\bibfnamefont {Y.}~\bibnamefont
  {Wu}}, \bibinfo {author} {\bibfnamefont {L.}~\bibnamefont {Yang}}, \bibinfo
  {author} {\bibfnamefont {Y.}~\bibnamefont {Zheng}}, \bibinfo {author}
  {\bibfnamefont {H.}~\bibnamefont {Deng}}, \bibinfo {author} {\bibfnamefont
  {Z.}~\bibnamefont {Yan}}, \bibinfo {author} {\bibfnamefont {Y.}~\bibnamefont
  {Zhao}}, \bibinfo {author} {\bibfnamefont {K.}~\bibnamefont {Huang}},
  \bibinfo {author} {\bibfnamefont {W.~J.}\ \bibnamefont {Munro}}, \bibinfo
  {author} {\bibfnamefont {K.}~\bibnamefont {Nemoto}}, \bibinfo {author}
  {\bibfnamefont {D.}~\bibnamefont {Zheng}}, \bibinfo {author} {\bibfnamefont
  {C.~P.}\ \bibnamefont {Sun}}, \bibinfo {author} {\bibfnamefont {Y.~X.}\
  \bibnamefont {Liu}}, \bibinfo {author} {\bibfnamefont {X.}~\bibnamefont
  {Zhu}}, \ and\ \bibinfo {author} {\bibfnamefont {L.}~\bibnamefont {Lu}},\
  }\bibfield  {title} {\enquote {\bibinfo {title} {An efficient and compact
  quantum switch for quantum circuits},}\ }\href@noop {} {\bibfield  {journal}
  {\bibinfo  {journal} {arXiv:1605.06747}\ } (\bibinfo {year}
  {2016})}\BibitemShut {NoStop}%
\bibitem [{\citenamefont {Caldwell}\ \emph {et~al.}(2017)\citenamefont
  {Caldwell} \emph {et~al.}}]{Caldwell2017}%
  \BibitemOpen
\bibfield  {author} {\bibinfo {author} {\bibfnamefont {S.}~\bibnamefont  {Caldwell}}, \bibfnamefont {N.}~\bibnamefont {Didier}, \bibfnamefont {C. A.}~\bibnamefont { Ryan}, \bibfnamefont {E. A.}~\bibnamefont { Sete}, \bibfnamefont {A.}~\bibnamefont { Hudson}, \bibfnamefont {P.}~\bibnamefont { Karalekas}, \bibfnamefont {R.}~\bibnamefont { Manenti}, \bibfnamefont {M. P. da}~\bibnamefont { Silva}, \bibfnamefont {R.}~\bibnamefont { Sinclair}, \bibfnamefont {E.}~\bibnamefont { Acala} \emph {et~al.}, \ } \bibfield  {title} {\enquote {\bibinfo {title} {{Parametrically Activated Entangling Gates Using Transmon Qubits}},}\
  }\href@noop {} {\bibfield  {journal} {\bibinfo  {journal} {Phys. Rev. Appl.}\
\textbf {\bibinfo {volume} {10}},\ \bibinfo {pages} {034050} } (\bibinfo {year} {2018})}\BibitemShut {NoStop}%
\bibitem [{\citenamefont {Reagor}\ \emph {et~al.}(2018)\citenamefont {Reagor}
  \emph {et~al.}}]{Reagor2018}%
  \BibitemOpen
\bibfield  {author} {\bibinfo {author} {\bibfnamefont {M.}~\bibnamefont {Reagor}}, \bibfnamefont {C. B.}\bibnamefont {Osborn}, \bibfnamefont {N.} \bibnamefont  {Tezak}, \bibfnamefont {A.} \bibnamefont {Staley}, \bibfnamefont {G.} \bibnamefont   {Prawiroatmodjo}, \bibfnamefont {M.} \bibnamefont  {Scheer}, \bibfnamefont {N.} \bibnamefont  {Alidoust}, \bibfnamefont {E. A.} \bibnamefont { Sete}, \bibfnamefont {N.} \bibnamefont   { Didier}, \bibfnamefont {M. P. da} \bibnamefont   { Silva} \emph {et~al.},\ }
\bibfield  {title} {\enquote {\bibinfo {title}
  {Demonstration of universal parametric entangling gates on a multi-qubit
  lattice},}\ }\href@noop {} {\bibfield  {journal} {\bibinfo  {journal} {Sci.
  Adv.}\ }\textbf {\bibinfo {volume} {4}},\ \bibinfo {pages} {eaao3603}
  (\bibinfo {year} {2018})}\BibitemShut {NoStop}%
\bibitem [{\citenamefont {Houck}\ \emph {et~al.}(2012)\citenamefont {Houck},
  \citenamefont {T\"ureci},\ and\ \citenamefont {Koch}}]{Houck2012}%
  \BibitemOpen
  \bibfield  {author} {\bibinfo {author} {\bibfnamefont {A.~A.}\ \bibnamefont
  {Houck}}, \bibinfo {author} {\bibfnamefont {H.~E.}\ \bibnamefont {T\"ureci}},
  \ and\ \bibinfo {author} {\bibfnamefont {J.}~\bibnamefont {Koch}},\
  }\bibfield  {title} {\enquote {\bibinfo {title} {On-chip quantum simulation
  with superconducting circuits},}\ }\href@noop {} {\bibfield  {journal}
  {\bibinfo  {journal} {Nat. Phys.}\ }\textbf {\bibinfo {volume} {8}},\
  \bibinfo {pages} {292} (\bibinfo {year} {2012})}\BibitemShut {NoStop}%
\bibitem [{\citenamefont {Georgescu}\ \emph {et~al.}(2014)\citenamefont
  {Georgescu}, \citenamefont {Ashhab},\ and\ \citenamefont
  {Nori}}]{Georgescu2014}%
  \BibitemOpen
  \bibfield  {author} {\bibinfo {author} {\bibfnamefont {I.~M.}\ \bibnamefont
  {Georgescu}}, \bibinfo {author} {\bibfnamefont {S.}~\bibnamefont {Ashhab}}, \
  and\ \bibinfo {author} {\bibfnamefont {F.}~\bibnamefont {Nori}},\ }\bibfield
  {title} {\enquote {\bibinfo {title} {Quantum simulation},}\ }\href@noop {}
  {\bibfield  {journal} {\bibinfo  {journal} {Rev. Mod. Phys.}\ }\textbf
  {\bibinfo {volume} {86}},\ \bibinfo {pages} {153} (\bibinfo {year}
  {2014})}\BibitemShut {NoStop}%
\bibitem [{\citenamefont {Kyriienko}\ and\ \citenamefont
  {S\o{}rensen}(2018)}]{Kyriienko2018}%
  \BibitemOpen
  \bibfield  {author} {\bibinfo {author} {\bibfnamefont {O.}~\bibnamefont
  {Kyriienko}}\ and\ \bibinfo {author} {\bibfnamefont {A.~S.}\ \bibnamefont
  {S\o{}rensen}},\ }\bibfield  {title} {\enquote {\bibinfo {title} {Floquet
  quantum simulation with superconducting qubits},}\ }\href@noop {} {\bibfield
  {journal} {\bibinfo  {journal} {Phys. Rev. Appl.}\ }\textbf {\bibinfo
  {volume} {9}},\ \bibinfo {pages} {064029} (\bibinfo {year}
  {2018})}\BibitemShut {NoStop}%
\bibitem [{\citenamefont {Clarke}\ and\ \citenamefont
  {Wilhelm}(2008)}]{Clarke2008}%
  \BibitemOpen
  \bibfield  {author} {\bibinfo {author} {\bibfnamefont {J.}~\bibnamefont
  {Clarke}}\ and\ \bibinfo {author} {\bibfnamefont {F.~K.}\ \bibnamefont
  {Wilhelm}},\ }\bibfield  {title} {\enquote {\bibinfo {title} {Superconducting
  quantum bits},}\ }\href@noop {} {\bibfield  {journal} {\bibinfo  {journal}
  {Nature}\ }\textbf {\bibinfo {volume} {453}},\ \bibinfo {pages} {1031}
  (\bibinfo {year} {2008})}\BibitemShut {NoStop}%
\bibitem [{\citenamefont {You}\ and\ \citenamefont {Nori}(2011)}]{You2011}%
  \BibitemOpen
  \bibfield  {author} {\bibinfo {author} {\bibfnamefont {J.~Q.}\ \bibnamefont
  {You}}\ and\ \bibinfo {author} {\bibfnamefont {F.}~\bibnamefont {Nori}},\
  }\bibfield  {title} {\enquote {\bibinfo {title} {Atomic physics and quantum
  optics using superconducting circuits},}\ }\href@noop {} {\bibfield
  {journal} {\bibinfo  {journal} {Nature}\ }\textbf {\bibinfo {volume} {474}},\
  \bibinfo {pages} {589} (\bibinfo {year} {2011})}\BibitemShut {NoStop}%
\bibitem [{\citenamefont {Devoret}\ and\ \citenamefont
  {Schoelkopf}(2013)}]{Devoret2013}%
  \BibitemOpen
  \bibfield  {author} {\bibinfo {author} {\bibfnamefont {M.~H.}\ \bibnamefont
  {Devoret}}\ and\ \bibinfo {author} {\bibfnamefont {R.~J.}\ \bibnamefont
  {Schoelkopf}},\ }\bibfield  {title} {\enquote {\bibinfo {title}
  {Superconducting circuits for quantum information: An outlook},}\ }\href@noop
  {} {\bibfield  {journal} {\bibinfo  {journal} {Science}\ }\textbf {\bibinfo
  {volume} {339}},\ \bibinfo {pages} {1169} (\bibinfo {year}
  {2013})}\BibitemShut {NoStop}%
\bibitem [{\citenamefont {Gu}\ \emph {et~al.}(2017)\citenamefont {Gu},
  \citenamefont {Kockum}, \citenamefont {Miranowicz}, \citenamefont {xi~Liu},\
  and\ \citenamefont {Nori}}]{GU2017}%
  \BibitemOpen
  \bibfield  {author} {\bibinfo {author} {\bibfnamefont {X.}~\bibnamefont
  {Gu}}, \bibinfo {author} {\bibfnamefont {A.~F.}\ \bibnamefont {Kockum}},
  \bibinfo {author} {\bibfnamefont {A.}~\bibnamefont {Miranowicz}}, \bibinfo
  {author} {\bibfnamefont {Y.-x.}~\bibnamefont {Liu}}, \ and\ \bibinfo {author}
  {\bibfnamefont {F.}~\bibnamefont {Nori}},\ }\bibfield  {title} {\enquote
  {\bibinfo {title} {Microwave photonics with superconducting quantum
  circuits},}\ }\href@noop {} {\bibfield  {journal} {\bibinfo  {journal}
  {Phys. Rep.}\ }\textbf {\bibinfo {volume} {718-719}},\ \bibinfo {pages}
  {1 } (\bibinfo {year} {2017})}\BibitemShut {NoStop}%
\bibitem [{\citenamefont {Barends}\ \emph {et~al.}(2013)\citenamefont
  {Barends}, \citenamefont {Kelly}, \citenamefont {Megrant}, \citenamefont
  {Sank}, \citenamefont {Jeffrey}, \citenamefont {Chen}, \citenamefont {Yin},
  \citenamefont {Chiaro}, \citenamefont {Mutus}, \citenamefont {Neill},
  \citenamefont {O'Malley}, \citenamefont {Roushan}, \citenamefont {Wenner},
  \citenamefont {White}, \citenamefont {Cleland},\ and\ \citenamefont
  {Martinis}}]{BarendsPRL2013}%
  \BibitemOpen
  \bibfield  {author} {\bibinfo {author} {\bibfnamefont {R.}~\bibnamefont
  {Barends}}, \bibinfo {author} {\bibfnamefont {J.}~\bibnamefont {Kelly}},
  \bibinfo {author} {\bibfnamefont {A.}~\bibnamefont {Megrant}}, \bibinfo
  {author} {\bibfnamefont {D.}~\bibnamefont {Sank}}, \bibinfo {author}
  {\bibfnamefont {E.}~\bibnamefont {Jeffrey}}, \bibinfo {author} {\bibfnamefont
  {Y.}~\bibnamefont {Chen}}, \bibinfo {author} {\bibfnamefont {Y.}~\bibnamefont
  {Yin}}, \bibinfo {author} {\bibfnamefont {B.}~\bibnamefont {Chiaro}},
  \bibinfo {author} {\bibfnamefont {J.}~\bibnamefont {Mutus}}, \bibinfo
  {author} {\bibfnamefont {C.}~\bibnamefont {Neill}}, \bibinfo {author}
  {\bibfnamefont {P.}~\bibnamefont {O'Malley}}, \bibinfo {author}
  {\bibfnamefont {P.}~\bibnamefont {Roushan}}, \bibinfo {author} {\bibfnamefont
  {J.}~\bibnamefont {Wenner}}, \bibinfo {author} {\bibfnamefont {T.~C.}\
  \bibnamefont {White}}, \bibinfo {author} {\bibfnamefont {A.~N.}\ \bibnamefont
  {Cleland}}, \ and\ \bibinfo {author} {\bibfnamefont {J.~M.}\ \bibnamefont
  {Martinis}},\ }\bibfield  {title} {\enquote {\bibinfo {title} {Coherent
  josephson qubit suitable for scalable quantum integrated circuits},}\
  }\href@noop {} {\bibfield  {journal} {\bibinfo  {journal} {Phys. Rev. Lett.}\
  }\textbf {\bibinfo {volume} {111}},\ \bibinfo {pages} {080502} (\bibinfo
  {year} {2013})}\BibitemShut {NoStop}%
\bibitem [{\citenamefont {Barends}\ \emph {et~al.}(2014)\citenamefont
  {Barends}, \citenamefont {Kelly}, \citenamefont {Megrant}, \citenamefont
  {Veitia}, \citenamefont {Sank}, \citenamefont {Jeffrey}, \citenamefont
  {White}, \citenamefont {Mutus}, \citenamefont {Fowler}, \citenamefont
  {Campbell}, \citenamefont {Chen}, \citenamefont {Chen}, \citenamefont
  {Chiaro}, \citenamefont {Dunsworth}, \citenamefont {Neill}, \citenamefont
  {O'Malley}, \citenamefont {Roushan}, \citenamefont {Vainsencher},
  \citenamefont {Wenner}, \citenamefont {Korotkov}, \citenamefont {Cleland},\
  and\ \citenamefont {Martinis}}]{Barends2014}%
  \BibitemOpen
  \bibfield  {author} {\bibinfo {author} {\bibfnamefont {R.}~\bibnamefont
  {Barends}}, \bibinfo {author} {\bibfnamefont {J.}~\bibnamefont {Kelly}},
  \bibinfo {author} {\bibfnamefont {A.}~\bibnamefont {Megrant}}, \bibinfo
  {author} {\bibfnamefont {A.}~\bibnamefont {Veitia}}, \bibinfo {author}
  {\bibfnamefont {D.}~\bibnamefont {Sank}}, \bibinfo {author} {\bibfnamefont
  {E.}~\bibnamefont {Jeffrey}}, \bibinfo {author} {\bibfnamefont {T.~C.}\
  \bibnamefont {White}}, \bibinfo {author} {\bibfnamefont {J.}~\bibnamefont
  {Mutus}}, \bibinfo {author} {\bibfnamefont {A.~G.}\ \bibnamefont {Fowler}},
  \bibinfo {author} {\bibfnamefont {B.}~\bibnamefont {Campbell}}
%\bibinfo   {author} {\bibfnamefont {Y.}~\bibnamefont {Chen}}, \bibinfo {author}  {\bibfnamefont {Z.}~\bibnamefont {Chen}}, \bibinfo {author} {\bibfnamefont  {B.}~\bibnamefont {Chiaro}}, \bibinfo {author} {\bibfnamefont  {A.}~\bibnamefont {Dunsworth}}, \bibinfo {author} {\bibfnamefont  {C.}~\bibnamefont {Neill}}, \bibinfo {author} {\bibfnamefont  {P.}~\bibnamefont {O'Malley}}, \bibinfo {author} {\bibfnamefont  {P.}~\bibnamefont {Roushan}}, \bibinfo {author} {\bibfnamefont  {A.}~\bibnamefont {Vainsencher}}, \bibinfo {author} {\bibfnamefont  {J.}~\bibnamefont {Wenner}}, \bibinfo {author} {\bibfnamefont {A.~N.}\  \bibnamefont {Korotkov}}, \bibinfo {author} {\bibfnamefont {A.~N.}\  \bibnamefont {Cleland}}, \ and\ \bibinfo {author} {\bibfnamefont {J.~M.}\  \bibnamefont {Martinis}},
\emph{et al.,}\ }\bibfield  {title} {\enquote {\bibinfo {title}
  {{Superconducting quantum circuits at the surface code threshold for fault
  tolerance}},}\ }\href@noop {} {\bibfield  {journal} {\bibinfo  {journal}
  {Nature}\ }\textbf {\bibinfo {volume} {508}},\ \bibinfo {pages} {500}
  (\bibinfo {year} {2014})}\BibitemShut {NoStop}%
\bibitem [{\citenamefont {Hatridge}\ \emph {et~al.}(2011)\citenamefont
  {Hatridge}, \citenamefont {Vijay}, \citenamefont {Slichter}, \citenamefont
  {Clarke},\ and\ \citenamefont {Siddiqi}}]{Hatridge}%
  \BibitemOpen
  \bibfield  {author} {\bibinfo {author} {\bibfnamefont {M.}~\bibnamefont
  {Hatridge}}, \bibinfo {author} {\bibfnamefont {R.}~\bibnamefont {Vijay}},
  \bibinfo {author} {\bibfnamefont {D.~H.}\ \bibnamefont {Slichter}}, \bibinfo
  {author} {\bibfnamefont {J.}~\bibnamefont {Clarke}}, \ and\ \bibinfo {author}
  {\bibfnamefont {I.}~\bibnamefont {Siddiqi}},\ }\bibfield  {title} {\enquote
  {\bibinfo {title} {Dispersive magnetometry with a quantum limited {SQUID}
  parametric amplifier},}\ }\href@noop {} {\bibfield  {journal} {\bibinfo
  {journal} {Phys. Rev. B}\ }\textbf {\bibinfo {volume} {83}},\ \bibinfo
  {pages} {134501} (\bibinfo {year} {2011})}\BibitemShut {NoStop}%
\bibitem [{\citenamefont {Roy}\ \emph {et~al.}(2015)\citenamefont {Roy},
  \citenamefont {Kundu}, \citenamefont {Chand}, \citenamefont {Vadiraj},
  \citenamefont {Ranadive}, \citenamefont {Nehra}, \citenamefont {Patankar},
  \citenamefont {Aumentado}, \citenamefont {Clerk},\ and\ \citenamefont
  {Vijay}}]{Roy2015}%
  \BibitemOpen
  \bibfield  {author} {\bibinfo {author} {\bibfnamefont {T.}~\bibnamefont
  {Roy}}, \bibinfo {author} {\bibfnamefont {S.}~\bibnamefont {Kundu}}, \bibinfo
  {author} {\bibfnamefont {M.}~\bibnamefont {Chand}}, \bibinfo {author}
  {\bibfnamefont {A.~M.}\ \bibnamefont {Vadiraj}}, \bibinfo {author}
  {\bibfnamefont {A.}~\bibnamefont {Ranadive}}, \bibinfo {author}
  {\bibfnamefont {N.}~\bibnamefont {Nehra}}, \bibinfo {author} {\bibfnamefont
  {M.~P.}\ \bibnamefont {Patankar}}, \bibinfo {author} {\bibfnamefont
  {J.}~\bibnamefont {Aumentado}}, \bibinfo {author} {\bibfnamefont {A.~A.}\
  \bibnamefont {Clerk}}, \ and\ \bibinfo {author} {\bibfnamefont
  {R.}~\bibnamefont {Vijay}},\ }\bibfield  {title} {\enquote {\bibinfo {title}
  {{Broadband parametric amplification with impedance engineering: Beyond the
  gain-bandwidth product}},}\ }\href@noop {} {\bibfield  {journal} {\bibinfo
  {journal} {Appl. Phys. Lett.}\ }\textbf {\bibinfo {volume} {107}},\ \bibinfo
  {pages} {262601} (\bibinfo {year} {2015})}\BibitemShut {NoStop}%
\bibitem [{\citenamefont {Kamal}\ \emph {et~al.}(2009)\citenamefont {Kamal},
  \citenamefont {Marblestone},\ and\ \citenamefont {Devoret}}]{Kamal}%
  \BibitemOpen
  \bibfield  {author} {\bibinfo {author} {\bibfnamefont {A.}~\bibnamefont
  {Kamal}}, \bibinfo {author} {\bibfnamefont {A.}~\bibnamefont {Marblestone}},
  \ and\ \bibinfo {author} {\bibfnamefont {M.~H.}\ \bibnamefont {Devoret}},\
  }\bibfield  {title} {\enquote {\bibinfo {title} {Signal-to-pump back action
  and self-oscillation in double-pump {J}osephson parametric amplifier},}\
  }\href@noop {} {\bibfield  {journal} {\bibinfo  {journal} {Phys. Rev. B}\
  }\textbf {\bibinfo {volume} {79}},\ \bibinfo {pages} {184301} (\bibinfo
  {year} {2009})}\BibitemShut {NoStop}%
\bibitem [{\citenamefont {Murch}\ \emph {et~al.}(2013)\citenamefont {Murch},
  \citenamefont {Weber}, \citenamefont {Macklin},\ and\ \citenamefont
  {Siddiqi}}]{Murch}%
  \BibitemOpen
  \bibfield  {author} {\bibinfo {author} {\bibfnamefont {K.~W.}\ \bibnamefont
  {Murch}}, \bibinfo {author} {\bibfnamefont {S.~J.}\ \bibnamefont {Weber}},
  \bibinfo {author} {\bibfnamefont {C.}~\bibnamefont {Macklin}}, \ and\
  \bibinfo {author} {\bibfnamefont {I.}~\bibnamefont {Siddiqi}},\ }\bibfield
  {title} {\enquote {\bibinfo {title} {Observing single quantum trajectories of
  a superconducting quantum bit},}\ }\href@noop {} {\bibfield  {journal}
  {\bibinfo  {journal} {Nature}\ }\textbf {\bibinfo {volume} {502}},\ \bibinfo
  {pages} {211} (\bibinfo {year} {2013})}\BibitemShut {NoStop}%
\bibitem [{\citenamefont {Didier}\ \emph {et~al.}(2018)\citenamefont {Didier},
  \citenamefont {Sete}, \citenamefont {da~Silva},\ and\ \citenamefont
  {Rigetti}}]{Didier2018}%
  \BibitemOpen
  \bibfield  {author} {\bibinfo {author} {\bibfnamefont {N.}~\bibnamefont
  {Didier}}, \bibinfo {author} {\bibfnamefont {E.~A.}\ \bibnamefont {Sete}},
  \bibinfo {author} {\bibfnamefont {M.~P.}\ \bibnamefont {da~Silva}}, \ and\
  \bibinfo {author} {\bibfnamefont {C.}~\bibnamefont {Rigetti}},\ }\bibfield
  {title} {\enquote {\bibinfo {title} {Analytical modeling of parametrically
  modulated transmon qubits},}\ }\href@noop {} {\bibfield  {journal} {\bibinfo
  {journal} {Phys. Rev. A}\ }\textbf {\bibinfo {volume} {97}},\ \bibinfo
  {pages} {022330} (\bibinfo {year} {2018})}\BibitemShut {NoStop}%
\bibitem [{\citenamefont {Chow}\ \emph {et~al.}(2009)\citenamefont {Chow},
  \citenamefont {Gambetta}, \citenamefont {Tornberg}, \citenamefont {Koch},
  \citenamefont {Bishop}, \citenamefont {Houck}, \citenamefont {Johnson},
  \citenamefont {Frunzio}, \citenamefont {Girvin},\ and\ \citenamefont
  {Schoelkopf}}]{Chow2009}%
  \BibitemOpen
  \bibfield  {author} {\bibinfo {author} {\bibfnamefont {J.~M.}\ \bibnamefont
  {Chow}}, \bibinfo {author} {\bibfnamefont {J.~M.}\ \bibnamefont {Gambetta}},
  \bibinfo {author} {\bibfnamefont {L.}~\bibnamefont {Tornberg}}, \bibinfo
  {author} {\bibfnamefont {J.}~\bibnamefont {Koch}}, \bibinfo {author}
  {\bibfnamefont {L.~S.}\ \bibnamefont {Bishop}}, \bibinfo {author}
  {\bibfnamefont {A.~A.}\ \bibnamefont {Houck}}, \bibinfo {author}
  {\bibfnamefont {B.~R.}\ \bibnamefont {Johnson}}, \bibinfo {author}
  {\bibfnamefont {L.}~\bibnamefont {Frunzio}}, \bibinfo {author} {\bibfnamefont
  {S.~M.}\ \bibnamefont {Girvin}}, \ and\ \bibinfo {author} {\bibfnamefont
  {R.~J.}\ \bibnamefont {Schoelkopf}},\ }\bibfield  {title} {\enquote {\bibinfo
  {title} {{Randomized Benchmarking and Process Tomography for Gate Errors in a
  Solid-State Qubit}},}\ }\href@noop {} {\bibfield  {journal} {\bibinfo
  {journal} {Phys. Rev. Lett.}\ }\textbf {\bibinfo {volume} {102}},\ \bibinfo
  {pages} {090502} (\bibinfo {year} {2009})}\BibitemShut {NoStop}%
\bibitem [{\citenamefont {Mei}\ \emph {et~al.}(2017)\citenamefont {Mei},
  \citenamefont {Chen}, \citenamefont {Tian}, \citenamefont {Zhu},\ and\
  \citenamefont {Jia}}]{Mei2017}%
  \BibitemOpen
  \bibfield  {author} {\bibinfo {author} {\bibfnamefont {F.}~\bibnamefont
  {Mei}}, \bibinfo {author} {\bibfnamefont {G.}~\bibnamefont {Chen}}, \bibinfo
  {author} {\bibfnamefont {L.}~\bibnamefont {Tian}}, \bibinfo {author}
  {\bibfnamefont {S.-L.}~\bibnamefont {Zhu}}, \ and\ \bibinfo {author}
  {\bibfnamefont {S.}~\bibnamefont {Jia}},\ }\bibfield  {title} {\enquote
  {\bibinfo {title} {Topologically protected quantum state transfer via edge
  states in superconducting circuits},}\ }\href@noop {} {\bibfield  {journal}
  {\bibinfo  {journal} {Phys. Rev. A} \ } \textbf {\bibinfo {volume} {98}},\
  \bibinfo {pages} {012331} (\bibinfo {year}
  {2018})}\BibitemShut {NoStop}%
\bibitem [{\citenamefont {Goldman}\ \emph {et~al.}(2014)\citenamefont
  {Goldman}, \citenamefont {Juzeli\=unas}, \citenamefont {\"Ohberg},\ and\
  \citenamefont {Spielman}}]{Goldman2014}%
  \BibitemOpen
  \bibfield  {author} {\bibinfo {author} {\bibfnamefont {N.}~\bibnamefont
  {Goldman}}, \bibinfo {author} {\bibfnamefont {G.}~\bibnamefont
  {Juzeli\=unas}}, \bibinfo {author} {\bibfnamefont {P.}~\bibnamefont
  {\"Ohberg}}, \ and\ \bibinfo {author} {\bibfnamefont {I.~B.}\ \bibnamefont
  {Spielman}},\ }\bibfield  {title} {\enquote {\bibinfo {title} {Light-induced
  gaugge fields for ultracold atoms},}\ }\href@noop {} {\bibfield  {journal}
  {\bibinfo  {journal} {Rep. Prog. Phys.}\ }\textbf {\bibinfo {volume} {77}},\
  \bibinfo {pages} {126401} (\bibinfo {year} {2014})}\BibitemShut {NoStop}%
\bibitem{Reedthesis} M. D. Reed, \emph{Entanglement and Quantum Error Correction with
Superconducting Qubits}, Ph.D. thesis, Yale University (2013).

\bibitem{Johnsonthesis} B. R. Johnson, \emph{Controlling Photons in Superconducting Electrical
Circuits}, Ph.D. thesis, Yale University (2010).

\bibitem{Kellythesis} J. S. Kelly, \emph{Fault-Tolerant Superconducting Qubits}, Ph.D. thesis,
University of California, Santa Barbara (2015).
\end{thebibliography}
\end{document}